\documentclass[iop,apj,twocolappendix,revtex4,numberedappendix,twocolumn]{emulateapj}

\usepackage{color}
\usepackage{mathtools}
\usepackage[normalem]{ulem}

\usepackage{times}
\usepackage{natbib}
\usepackage{graphicx}
\usepackage{amsmath}
\usepackage{color}
\newcommand{\str}{Str\"{o}mgren}

\newcommand{\nH}{n_{\infty}}

\newcommand{\hi}{\rm H\,{\textsc i}}
\newcommand{\hii}{\rm H\,{\textsc{ii}}}
\newcommand{\myscale}{1.15}
\newcommand{\msun}{M_\sun}

\newcommand{\mbh}{M_{\rm BH}}

\newcommand{\rbondi}{r_{\rm B}}
\newcommand{\magvolt}{|\vec{\omega}|}

\newcommand{\revI}[1]{#1}
\newcommand{\revII}[1]{#1}
\newcommand{\revIII}[1]{#1}

\shorttitle{Radiation-driven turbulent accretion onto massive black holes}
\shortauthors{Park et al.}
\begin{document}
\title{Radiation-driven turbulent accretion onto massive black holes}

%\author{KwangHo Park}
\author{KwangHo Park{\altaffilmark{1}}, John H. Wise{\altaffilmark{1}}, and
Tamara Bogdanovi\'c{\altaffilmark{1}}} \email{kwangho.park@physics.gatech.edu}

\affil{{\altaffilmark{1}}Center for Relativistic Astrophysics, School of Physics,
Georgia Institute of Technology, Atlanta, GA 30332, USA; \\ kwangho.park@physics.gatech.edu}

\begin{abstract} 

Accretion of gas and interaction of matter and radiation are at the
heart of many questions pertaining to black hole (BH) growth and
coevolution of massive BHs and their host galaxies. 
To answer them it is critical to quantify how the ionizing radiation
that emanates from the innermost regions of the BH accretion flow
couples to the surrounding medium and how it regulates the BH fueling.
In this work we use high resolution 3-dimensional (3D)
radiation-hydrodynamic simulations with the code {\it Enzo}, equipped
with adaptive ray tracing module {\it Moray}, to investigate
radiation-regulated BH accretion of cold gas. Our simulations
reproduce findings from an earlier generation of 1D/2D simulations:
the accretion powered UV and X-ray radiation forms a highly ionized
bubble, which leads to suppression of BH accretion rate characterized
by quasi-periodic outbursts. A new feature revealed by the 3D
simulations is the highly turbulent nature of the gas flow in
vicinity of the ionization front. \revII{During quiescent periods
between accretion outbursts, the ionized bubble shrinks in size} 
\revII{and the gas density
that precedes the ionization front increases. Consequently,} the 3D simulations
show oscillations in the accretion rate of only $\sim$\,2-3 orders
of magnitude, significantly smaller than 1D/2D models. We calculate
the energy budget of the gas flow and find that turbulence is the
main contributor to the kinetic energy of the gas but corresponds
to less than 10\% of its thermal energy \revII{and thus} does not
contribute significantly to the pressure support of the gas.

\end{abstract}


%%%%%%%%%%%%%%%%%%%%%%%%%
\section{Introduction}

The existence of supermassive black holes (BHs) observed as quasars at
high redshift challenges our understanding of the formation of seed BHs
and their growth history in the early Universe \citep{Fan:2001,
Fan:2003, Fan:2006, Willott:2003, Willott:2010, Wu:2015}. Several
scenarios have been suggested for the origin of seed BHs in mass range
of $10^2$--$10^5\,\msun$ (intermediate-mass black holes; IMBHs)
such as Population~III remnants \citep{BrommCL:99,AbelBN:00,MadauR:01},
collapse of primordial stellar clusters
\citep{Devecchi:2009,Davies:2011,Katz:2015}, and direct collapse of
pristine gas \citep{Carr:84,HaehneltNR:98,Fryer:01,BegelmanVR:06,
ChoiSB:13,YueFSXC:14,Regan:2017}. Alternatively, mildly metal-enriched
gas can gravitationally collapse to form an IMBH in pre-galactic disk
\citep{LodatoN:2006, OmukaiSH:08}. However, even such massive seed BHs
still have to go through rapid growth to be observed as billion solar
mass quasars at $z\sim7$. Thus, a realistic estimate of accretion rate
can provide a test to plausible theoretical scenarios \citep{MadauR:01,
VolonteriHM:03,YooM:04,Volonteri:05}.

The most perplexing issue related to the existence of high redshift
quasars is the ubiquity of radiative feedback associated with the
central BHs which should, in principle, preclude the rapid growth.
Indeed, several works have shown that the radiative feedback from BHs
regulates the gas supply from large spatial scales, slowing down the
growth of BHs significantly, despite of the availability of high density
neutral cold gas in the early Universe
\citep{AlvarezWA:09,MiloCB:09,ParkR:11, ParkR:12, ParkR:13, ParkRDR:14a,
ParkRDR:14b, Park:2016}. For example, the radiation has been shown to
readily suppress accretion in the regime when $\mbh \nH \la 10^9\,\msun
{\rm cm}^{-3}$ \citep{ParkR:12}, where $\mbh$ and $\nH$ are the BH mass
and gas number density unaffected by the BH feedback in units of $\msun$
and ${\rm cm}^{-3}$, respectively. Self-regulation occurs when the
ionizing radiation from the BH accretion disk produces a hot and
rarefied bubble around the BH. Simulations show that the average
accretion rate is suppressed to \revI{a few} percent of the classical Bondi
accretion rate (i.e., measured in absence of radiative feedback) and
that the accretion rate shows an oscillatory behavior due to the
accretion/feedback loop. As $\mbh$ or $\nH$ increases, the
 $\mbh \nH \ga 10^9\,\msun {\rm cm}^{-3}$ threshold is crossed. In this regime accretion onto the BH makes a critical transition to a high accretion rate regime so-called {\it hyper-accretion} where the radiation pressure cannot longer resist the gravity of the inflowing gas
\citep{Begelman:79, PacucciVF:2015, ParkRDR:14a,Inayoshi:2016,Park:2016}.

The local 1D and 2D simulations of accretion mediated by radiative
feedback have been extensively used to explore these accretion regimes
\citep{ParkR:11, ParkR:12, ParkRDR:14b}. They commonly adopt an
assumption of spherical symmetry of the accretion flow and isotropy of
ionizing radiation emerging from a radiatively efficient accretion disk
\citep{ShakuraS:73}. Specifically, 1D simulations have been used to
efficiently explore the parameter space of radiative efficiency, BH
mass, gas density/temperature, and spectral index of the radiation. In
2D simulations, an additional degree of freedom revealed the growth of
the Rayleigh--Taylor instability across the ionization front which is
suppressed and kept in check by ionizing radiation \citep{ParkRDR:14a}.
2D simulations are also useful to study the effect of anisotropic
radiation from an accretion disc that preferentially produces radiation
perpendicular to the disk plane \citep[e.g.,][]{Sugimura:2016}. 

Although the assumption of spherical symmetry is reasonable in the setup
where an isolated BH is accreting from a large scale reservoir of
uniform and neutral gas, it has been long known that certain physical
processes can only be reliably captured in full 3D simulations. A well
known example is a spherical accretion shock instability (SASI), found
in supernovae simulations, where an extra degree of freedom in 3D
simulations is found to significantly affect the dynamics of gas and
explosion of a supernova \citep[e.g.,][]{BlondinS:07}.  Furthermore, 3D
simulations of radiation-regulated accretion are necessary in order to
capture fueling and feedback of BHs in a more complex cosmological
context. 

The main aim of this paper is to extend numerical studies of accretion
mediated by radiative feedback to full 3D local simulations and identify
any physical processes that have not been captured by the
local 1D/2D simulations. In order to achieve this we carry out a suite
of high resolution 3D hydrodynamic simulations with the adaptive mesh
refinement (AMR) code {\it Enzo}, equipped with the adaptive ray tracing
module {\it Moray} \citep{Wise:2011,Bryan:2014}. The results to be
presented in the next sections corroborate the role of ionizing
radiation in regulating accretion flow which causes an oscillatory
behavior of gas accretion. More interestingly, our simulations capture the development of the radiation-driven turbulence, which plays a role in the BH fueling during periods of quiescence. In
Section~\ref{sec:method}, we explain the basic accretion physics and the
numerical procedure. We present the results in Section~\ref{sec:results}
and discuss and summarize them in Section~\ref{sec:discussion}.

%%% SECTION 2 %%%

\section{Methodology}
\label{sec:method}
\subsection{Calculation of Accretion Rate}

One of the first steps in quantifying the radiation-regulated accretion
onto BHs starts with the estimate of accretion rate. We briefly review
two main approaches that have been used as a part of different numerical
schemes in earlier works.  

A straightforward approach to accretion rate measurement in simulations
is to {\it read} the mass flux directly through a spherical surface
centered around the BH. This approach is usually used in simulations
that employ spherical polar coordinate systems and it requires that the
sonic point in the accretion flow is resolved in order to return
reliable results \citep[e.g.,][]{NovakOC:11, ParkR:11}. For example, a
minimum radius of the computation domain $r_{\rm min} \sim
10^{-3}\,(\mbh/10^4\,\msun)$\,pc has been used in some of our earlier works, where logarithmically spaced radial grids make it possible to achieve a high spacial resolution close to the BH \citep{ParkR:11,
ParkR:12,ParkR:13, Park:2016, Park:2017}. 
%The size of the highest resolution element in radial direction corresponds to only a few percent of the minimum radius, where the gas is assumed to be accreted to the BH, and thus the sonic radius is resolved. 

Another way to estimate the accretion rate is to {\it infer} the
accretion rate from gas properties near the BH assuming a spherically
symmetric accretion onto a point source \citep{Bondi:52}. In this case,
it is important to resolve the Bondi radius within which the
gravitational potential by the BH dominates over the thermal energy of
the gas
\begin{equation} r_{\rm B}= \frac{G\mbh}{c_{s,\infty}^2}. \end{equation}
Here, $c_{s,\infty}$ is the sound speed of the gas far from the BH. For
the gas with temperature $T_\infty=10^4$\,K, $r_{\rm B} \sim
0.5\,(\mbh/10^4\,\msun)$\,pc, which is about 2 orders of magnitude
larger than $r_{\rm min}$ used in the \revI{aforementioned} mass flux 
measurement method. This
approach consequently imposes less stringent requirements on numerical
resolution and is often used in simulations that employ Cartesian
coordinate grids.

The classical Bondi accretion rate can then be estimated in terms
of the properties of the gas and BH mass
\begin{equation} 
\begin{split} \dot{M}_{\rm B}  =& \; 4 \pi \lambda_{\rm B}
\rho_\infty \frac{G^2 \mbh^2}{c_{\rm
s,\infty}^{3}} \\
=& \; 8.7\!\times\!10^{-4}\!
\left(\frac{n_\infty}{10^3\,{\rm cm^{-3}}} \right)
\left(\frac{T_\infty}{10^4\,{\rm K}} \right)^{-\frac{3}{2}}\\
& \times \left(\frac{\mbh}{10^4 M_\odot} \right)^2\, \msun\,{\rm yr}^{-1}.
\label{eq:M_B}
\end{split} \end{equation}
where $\lambda_{\rm B}$ is a dimensionless parameter ranging from
$1/4$ for adiabatic gas ($\gamma=5/3$) to $e^{3/2}/4$ for isothermal
gas ($\gamma=1$) \revI{which we adopt to represent the value for
the classical Bondi rate for the purpose of normalization}.

The latter approach is commonly preferred in large cosmological
simulations where a large dynamic range of spatial scales is involved
and BHs are often treated as sink particles. The Eddington-limited
Bondi rate is the most common recipe for
the growth of BHs  used in the literature
\begin{equation}
\dot{M}_{\rm BH}=
{\rm min} \left[\dot{M_{\rm B}}, \frac{L_{\rm Edd}}{\eta c^2}\right] 
\label{eq_Mdot}
\end{equation}
where $\eta$ is the radiative efficiency and $c$ is the speed of light.
In this approach the Eddington rate is the maximum accretion rate
onto the BH set by the radiative feedback from the BH and defined as 
\begin{equation}
L_{\rm Edd} = \frac{4\pi G\mbh m_{\rm p}c}{\sigma_{\rm T}} \simeq
1.26\!\times\!10^{38}\left(\frac{\mbh}{\msun}\right)\,{\rm erg\ s}^{-1}
\end{equation}
for pure hydrogen gas, where $m_{\rm p}$ is the proton mass and $\sigma_{\rm
T}$ is the Thomson cross section. From the accretion rate in
Equation~(\ref{eq_Mdot}), the accretion luminosity is calculated as
\begin{equation} 
L_{\rm acc}= \eta \dot{M}_{\rm BH} c^2 
\end{equation} 
where we adopt a constant radiative efficiency $\eta = 0.1$ assuming a thin
disk model \citep{ShakuraS:73}.  

%%% TABLE1 %%%%%%%%%%%%%%%%%%%%%%%%%%
\begin{table*}[thb]
\begin{center}
\caption{Simulation Parameters}
\begin{tabular}{lccccccc}
\hline 
\hline
   & $\mbh$ & $\nH$ & $L_{\rm box}$ & Top & $\Delta L_{\rm max}$ & $\Delta L_{\rm min}$  & $N_\nu$\\
Run ID & $(\msun)$	&  $({\rm cm}^{-3})$ & (kpc) &  Grids & (pc) & (pc)  &   \\
\hline
{M4N3} & $10^4$  & $10^3$ & 0.04 & $32^3$  & 1.25 & 0.156 & 4   \\
%{M4N3mod} & $10^4$  & $10^3$ & 0.04 & $32^3$  & 3    \\
{M4N3sec} & $10^4$  & $10^3$ & 0.04 & $32^3$  & 1.25 & 0.156  & 4   \\
{M4N3E8mod} & $10^4$  & $10^3$ & 0.04 & $32^3$  & 1.25 & 0.156  &8  \\
\revI{M4N3E8R64mod} & $10^4$  & $10^3$ & 0.04 & $64^3$  & 0.625 & 0.078  &8  \\
%M4N3h & $10^4$  & $10^3$ & 0.04 & $64^3$ & 3     \\
{M4N4} & $10^4$   & $10^4$ & 0.02 & $32^3$  & 0.625 & 0.078 & 4    \\
{M6N1} & $10^6$  & $10^1$ & 4.0 & $32^3$  & 125  & 15.6 & 4   \\
%M6N1h & $10^6$  & $10^1$ & 4.0 & $64^3$  & 3    \\
%M6N1l & $10^6$  & $10^1$ & 4.0 & $32^3$  & 2   \\
%M6N1E8l & $10^6$  & $10^1$ & 4.0 & $32^3$  & 3   \\
%M6N1E8T6  & $10^6$  & $10^1$ & 4.0 & $64^3$  & 3  \\
%M6N1E8T6l & $10^6$  & $10^1$ & 4.0 & $64^3$  & 2  \\
%M2N5 & $10^2$  & $10^5$ & 4.0 & $32^3$  & 3  &  4 &   \\
\hline
\end{tabular}
\label{table:para}
\end{center}
\end{table*}

We adopt the latter approach (i.e., the Bondi prescription) for
calculating the BH accretion rate, which is well suited to the numerical
scheme used in this work based on Cartesian coordinate grid. Note
however that instead of the properties of neutral gas we use $\rho_{\rm
HII}$ and $c_{s,{\rm HII}}$, which denote the density and sound speed of
the gas under the influence of BH radiation. The BH accretion rate is
{\it estimated}  \citep{Kim:2011}
\begin{equation} \dot{M}_{\rm BH}  =  4 \pi \rho_{\rm HII}
\frac{G^2 \mbh^2}{c_{\rm s,HII}^{3}}, \label{eq:mdot_hii} \end{equation}
\revI{where we assume $\lambda_{\rm B} \sim 1$. Note however that
in reality the equation of state of the gas changes depending on
the efficiency of cooling and heating.} Therefore, we assume that
gas accretion onto the BH inside the \str~radius still occurs in a
manner similar to Bondi accretion. The implication is that the BH
does not accrete cold gas ($T \sim 10^4$\,K) directly from larger
spatial scales, but is instead fueled by the ionized gas heated by
its own radiation. The temperature of this photo-heated gas is $T
\sim 4\times 10^4$\,K for the spectral index of ionizing radiation
$\alpha=1.5$, assuming pure hydrogen gas \citep[see][]{ParkR:11,ParkR:12}.
Note that the mean accretion rate can be analytically derived from
the pressure equilibrium in time-averaged gas density and temperature
profiles between the neutral and ionized region, such that $\rho_\infty
T_\infty \simeq \revI{2} \rho_{\rm HII} T_{\rm HII}$ \citep{ParkR:11}.
Equation~(\ref{eq:mdot_hii}) can be rewritten to estimate the mean
accretion rate as a function of ($T_\infty/T_{\rm HII}$) as
\begin{equation} \langle\dot{M}_{\rm BH}\rangle \simeq
\left[\frac{T_\infty}{T_{\rm HII}}\right]^{5/2} \dot{M}_{\rm B}
\label{eq:mdot_mean} \end{equation}
Because $\rho_{\rm HII}$  and $c_{\rm s,HII}$ evolve constantly with
time, so does the new Bondi radius for the photo-heated gas $\rbondi ' =
G\mbh /c_{\rm s,HII}^2 \propto \mbh/T_{\rm HII}$. Note that we aim to
resolve the new Bondi radius $\rbondi '= \rbondi (T_\infty/T_{\rm
HII})$. For the photo-heated hydrogen gas $T_{\rm HII} \sim 4\times 10^4$\,K and $\rbondi ' = 0.13\,(\mbh/10^4\,\msun)$\,pc which is comparable to or larger than the numerical resolution achieved in vicinity of the BH in this work (represented by $\Delta L_{\rm min}$ in
Table~\ref{table:para}). 

%%% FIG1 %%%

\begin{figure*}[t]
\epsscale{1.2} \plotone{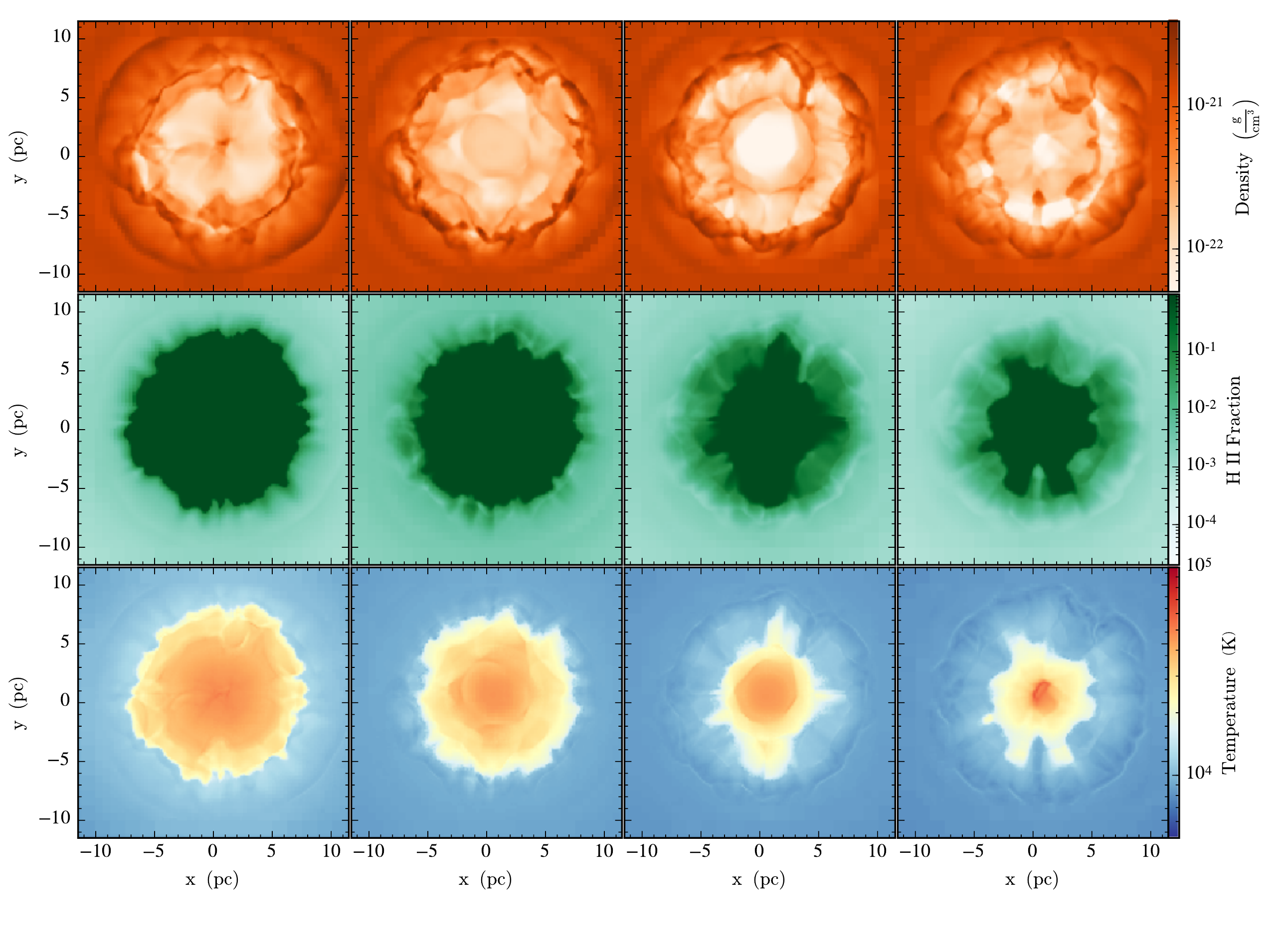} 

\caption{Slices of \revI{M4N3E8mod at $t=5.75$ (burst of
accretion), 5.82, 5.88, and 5.95\,Myr} from left to right panels.
The top, middle, and bottom panels show gas density, \hii~fraction,
and gas temperature, respectively. Due to the inhomogeneity of the gas,
the \str~radius is
approximately spherically symmetric but with a large variation in
distances from the BH. At peak accretion rate the \str~radius reaches its
maximum size, which can also be seen in the temperature slices.
As the accretion rate decreases (from left to right), the
\hii~region shrinks but never vanishes.}
\label{fig:snapshot1} \end{figure*}

%%% section22

\subsection{Radiation-Hydrodynamic Simulations with Enzo+Moray}

We perform local 3D radiation-hydrodynamic simulations in Cartesian
coordinates using the AMR code {\it Enzo} coupled with the adaptive ray
tracing module {\it Moray} \citep{Wise:2011, Bryan:2014}.

The BH is modeled as a sink particle and is located at the center of the
computation domain. We fix the position of the BH throughout the
simulation and update the BH mass to account for growth through
accretion. We assume a simple initial condition of uniform
temperature and density, as well as zero velocity and metalicity for the background gas \citep[see][for dusty accretion onto BHs]{Yajima:2017}. \revI{We assume an adiabatic gas ($\gamma=5/3$) and} 
neglect its self-gravity. We select
$\mbh=10^4\,\msun$, $\nH=10^{3}\,{\rm cm}^{-3}$, and $T_\infty =10^4$\,K
as the baseline setup in the feedback-dominated regime \citep{ParkR:12}.  Note
that the combination of $\mbh$ and $\nH$ can be extended to other sets
of simulations which return \revI{qualitatively comparable} results (i.e., Equation~(\ref{eq:mdot_mean}) holds and the size of \hii~region is $\propto \mbh$) when $\mbh\, \nH$ is kept constant.
Table~\ref{table:para} lists the parameters for each simulation. We
use the following naming convention for simulations: runs are denoted as
`MmNn' where the BH mass is $\mbh=10^{m}\,\msun$ and gas number density
is $\nH=10^{n}\,{\rm cm}^{-3}$.
 
We use numerical resolution of $32^3$ on the top grid in most of our simulations, except where noted otherwise. This resolution corresponds to a coarsest resolution element with the size, $\Delta L_{\rm max}$, 
listed in Table~\ref{table:para}. With 3 levels of refinement our simulations attain the finest resolution of $\Delta L_{\rm min}= 0.156
(\mbh/10^4\,\msun)\,$pc. The strategy we adopt for AMR is to achieve
the highest level of resolution both in the central region near the BH
and around the ionization front. In order to accomplish this, we
enforce the highest level of refinement within the box of a size $\Delta
L_{\rm max}$ 
%\jhw{[Surely, the innermost box is larger than a root grid
%cell width.  Double check this.]} 
centered around the BH (see
Table~\ref{table:para}). In addition to this requirement we also use
the local gradients of all variables (i.e., density, energy, pressure,
velocity, and etc) to flag cells for refinement around the ionization
front. Outflowing boundary conditions are imposed on all boundaries to
prevent reflection of density waves, which form due to the expansion
of the ionized region.

%\tb{[We need the discussion of heating and cooling processes and any
%other relevant physics bits that are used in simulations.  Also discuss
%any relevant criteria for the time step: is it set by hydrodynamics only
%or does it also account for the cooling rate or else. Then continue with
%the discussion of {\it Moray} in the next paragraph.]}

We use the non-equilibrium chemistry model for $\hi$, $\hii$, ${\rm
He}\,\textsc{i}$, ${\rm He}\,\textsc{ii}$, ${\rm He}\,{\textsc{iii}}$,
and $e^{-}$ implemented in {\it Enzo} \citep{Abel:1997, Anninos:1997}. 
For simplicity, we consider the photo-heating and cooling of pure hydrogen 
gas and neglect the effects of radiation pressure, which are relatively minor. The radiation pressure on both the electron gas and neutral hydrogen is weaker than the local gravity by the BH and negligible relative to the thermal pressure which plays a central role in creating outflows inside the \hii~region \citep[see][for details]{ParkR:12}. The photo-heating of the hydrogen-helium gas mixture is known to return a higher temperature inside
the \hii~region (i.e., $T_{\rm HII} \sim 6\times 10^4$\,K), which
leads to somewhat lower accretion rate and qualitatively similar results
to the pure hydrogen gas. The Compton heating plays a minor role
compared to the photo-heating when the radiation spectrum is soft
\citep{ParkRDR:14b} and is therefore neglected.

%\tb{[The sentences above are confusing - clarify. Are you saying
%that you are considering the heating of the pure hydrogen gas but
%cooling from the H-He gas?]} 
% KP: Just heating and cooling by H not by He. However, the chemistry
% evolved. I will keep it as it is.      

%\tb{[Elaborate, why is it a minor effect. Also we need to introduce
%the
%concept of secondary ionizations, since this is mentioned in the text.]}  
%\jhw{[We should include some details about secondary ionizations since
%it's used in one model.]}
% KP: In section 3.2. simple descripton of secondary ionization was
% introduced

Radiation emitted from the innermost parts of the BH accretion flow is
propagated through the computational domain by the module {\it Moray},
which solves radiative transfer equation coupled with equations of
hydrodynamics. Specifically, {\it Moray} accounts for the
photo-ionization of gas and calculates the amount of photo-heating,
which contributes to the total thermal energy of the gas.  {\it Moray}
uses adaptive ray tracing \citep{AbelW:02} that is based on the HEALPix
framework \citep{Gorski:05}. In this approach the BH is modeled as a
radiation point source that emits {$12\times 4^3$} rays which are then split
into four child rays so that a single cell is sampled by at least 5.1
rays. The adaptive ray splitting occurs automatically when the solid
angle associated with a single ray increases with radius or if the ray
encounters a high resolution AMR grid.  The radiation field is updated
in every time step that corresponds to the grid with the finest
numerical resolution as set by the Courant-Friedrichs-Lewy condition.

%%% SED %%%
\subsection{Modified power-law spectral energy distribution}

We describe the accretion luminosity with a power law spectrum, $L_\nu
=C \nu^{-\alpha}$, where $\alpha$ is the spectral index and $C$ is the
normalization constant. The frequency integrated (bolometric) luminosity
between the energy $h\nu_1$ and $h\nu_2$ is $L_{\rm acc} =
C(\nu_2^{1-\alpha} - \nu_1^{1-\alpha})/(1-\alpha)$ for $\alpha \ne 1$.
{\it Moray} models the SED with $N_\nu = 4-8$ discrete energy bins that
are equidistant in log-space between $E_{\rm min}=h\nu_{\rm min} =
13.6$\,eV and $E_{\rm max}=h\nu_{\rm max}=100$\,keV. A fraction of
energy is allotted to each bin assuming the spectral index $\alpha=1.5$.
The mean photon energy of the $n$-th bin $\langle E_n \rangle$ between
energies $E_{n-1}$ and $E_{n}$ is calculated as
\begin{equation} \langle{E}_n\rangle =\frac{\alpha}{\alpha
-1}\frac{(E_n^{1-\alpha}-E_{n-1}^{1-\alpha})}
{(E_n^{-\alpha}-E_{n-1}^{-\alpha})}\,\, {\rm for}\,\, \alpha > 1
\end{equation}
which returns the mean energy of $\langle E \rangle \simeq 40.3$\,eV for
the entire energy range. Table~\ref{table:sed} lists the mean energy
($h\nu_n$) and the dimensionless fraction of bolometric luminosity allotted to each bin ($L_n$). The number of ionizing photons in each bin is then
calculated as $L_n L_{\rm acc}/\langle{E}_n\rangle$.

The prescription for modeling the spectrum outlined above ensures that
the total power of emitted radiation is divided among the chosen
number of energy bins to model the SED. In addition to the energy
spectrum of radiation, another crucial consideration for radiation
transport is the number of ionizing photons contained in each energy
bin, as given by the photon spectrum. In order to maintain consistency
in terms of the photon spectrum among simulations with different values
of $N_\nu$, we introduce a modification to the number of photons in the
first energy bin. This ensures that the number of near-UV ionizing photons, which are responsible for the bulk of photo-ionizations and
photo-heating, remains approximately the same regardless of which SED
model is used. Specifically, the mean energy of the lowest energy
bin determines the temperature of \hii~region, which is why we adopt the
same energy of $E_1=28.4$\,eV for 4 and 8-bin SED models (the original
first entry for $N_\nu=8$ is $E_1 = 21.5$\,eV). The fraction of energy
in this bin is accordingly adjusted from 0.4318 to 0.5704 so to preserve
the same number of ionizing photons. One consequence of this
optimization performed simultaneously in terms of the luminosity and
number of ionizing photons is that the sum of luminosity fractions
allotted to all energy bins ($L_n$) is no longer exactly equal to 100\%,
as can be seen from Table~\ref{table:sed}. 
\revI{We choose the 4-bin model without modification 
as a reference since it still returns a consistent result with
\citet{ParkR:11} using the minimum number of energy bins.} 

In the Appendix we include a
detailed discussion of the SED optimization used in this work, along
with convergence tests, and a comparison with SED models published in
\citet{Mirocha:2012}. We explore a full set of modeled SEDs in our
simulations and mark those that are modified as outlined in this
section in Table~\ref{table:para}. For example, `E8mod' indicates
the case of modified SED with 8 energy bins and `sec' marks the
case with secondary ionizations included \revI{to consider the
effect by energetic particles produced by high energy X-ray photons}.

%\tb{[Secondary ionizations mentioned for the first time and without explanation. Elaborate here or where you describe chemistry and photo-heating.]}. 

%\tb{[I would omit this paragraph]} The configuration for $(h\nu_n, I_n)$ for different $N_\nu$ in
%Table~\ref{table:sed} is consistent in terms of the mean energy of
%photons and the number of ionizing photons.  Note that this configuration 
%is different from the optimized SED 
%by \citet{Mirocha:2012} where $10^2$--$10^4$\,eV is selected for
%X-ray sources since the role of UV photons are important in our cases 
%when the spectrum is soft. We also test a
%set of optimized SED for $13.6$--$10^5$\,eV with $\alpha=1.5$ using
%the same method in \citet{Mirocha:2012} to which we compare our
%main results in the Appendix.

%%% TABLE1 %%%%%%%%%%%%%%%%%%%%%%%%%%
\begin{table}[t]
\begin{center}
\caption{SEDs for Power-law BH radiation with $\alpha=1.5$}
\begin{tabular}{ccc}
\hline 
\hline
%  		& $n_1$		& $n_2$		& $n_3$		& $n_4$	 \\	
%$h\nu$\,(eV) 	& 28.4 		& 263.0 	& 2435.3 	& 22551.1 \\	
%$I_{\nu}$ 	&0.6793		& 0.2232	& 0.0734	& 0.0241 \\	

$n_\nu$    &  $N_\nu=4$	& $N_\nu=8$ \\	
\hline
1 & (28.4, 0.6793) & (28.4, 0.5704) \\
2 & (263.0, 0.2232) & (65.3, 0.2475) \\ 
3 & (2435.3, 0.0734) & (198.7, 0.0813) \\
4 & (22551.1, 0.0241) & (604.5, 0.0813) \\
5 & ... & (1839.5, 0.0466) \\
6 & ... & (5597.8, 0.0267) \\
7 & ... & (17034.3, 0.0153) \\
8 & ... & (51836.1, 0.0088) \\
\hline
\end{tabular}
\tablecomments{Each entry is given as ($h\nu_n, L_n$) where $h\nu_n$
is the photon energy in units of eV and $L_n$ is the dimensionless fraction of bolometric luminosity assigned to each energy bin, $n_\nu$. See text for details.} 
\label{table:sed}
\end{center}
\end{table}

%%%%%%%%%%%%%%%%%%%%%%%%%%%%%%%%%%%%%%%%%%%%%%%%%%

\section{Results}
\label{sec:results}

Our 3D simulations are characterized by the oscillatory behavior of the
accretion rate and size of ionized region, which is analogous to the
previous 1D/2D simulations. Our simulations also show the presence of the
turbulent gas motion driven by the fluctuating ionized region, which has not been captured in lower dimensionality simulations.

\subsection{Formation of ionized region and oscillatory behavior} 

%%%% FIG2 %%%%%%%%%%%%%%%%%%%%%%%%%%%%%%%%%%%%%%%%%%

\begin{figure}[t] \epsscale{\myscale} 
\plotone{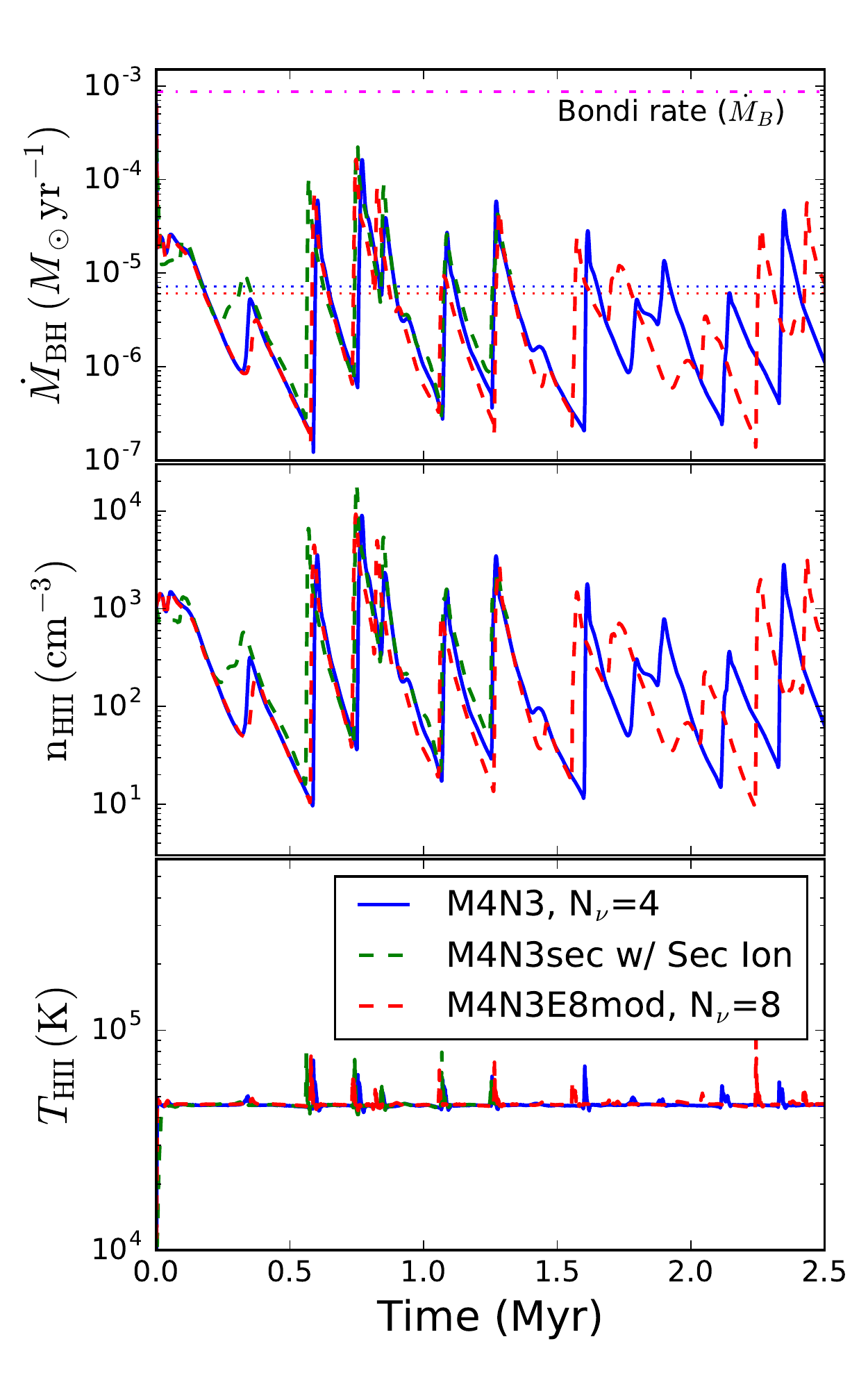} 

\caption{From top to bottom: accretion rate onto the BH (with mean values
shown as dotted lines), density and temperature inside the \hii~region near the BH for the runs M4N3 (blue solid), M4N3sec (green dashed), and M4N3E8mod (red dashed). The accretion rate exhibits oscillatory behavior and spans
$\sim$\,3 orders of magnitude between the maximum ($\sim
10^{-4}\,{\rm M}_\odot/{\rm yr}$) and minimum ($\sim 10^{-7}\,{\rm
M}_\odot/{\rm yr}$). 
\revI{Dot-dashed line shows the Bondi rate ($\dot{M}_{\rm B}$) for neutral gas with $T_\infty=10^4$\,K.}}

\label{fig:acc_rate} \end{figure}

%%% FIG3 %%%%%%%%%%%%%%%%%%%%%%%%%%%%%%%%%%%%%%%%%%

\begin{figure*}[t] \epsscale{1.16}
\plottwo{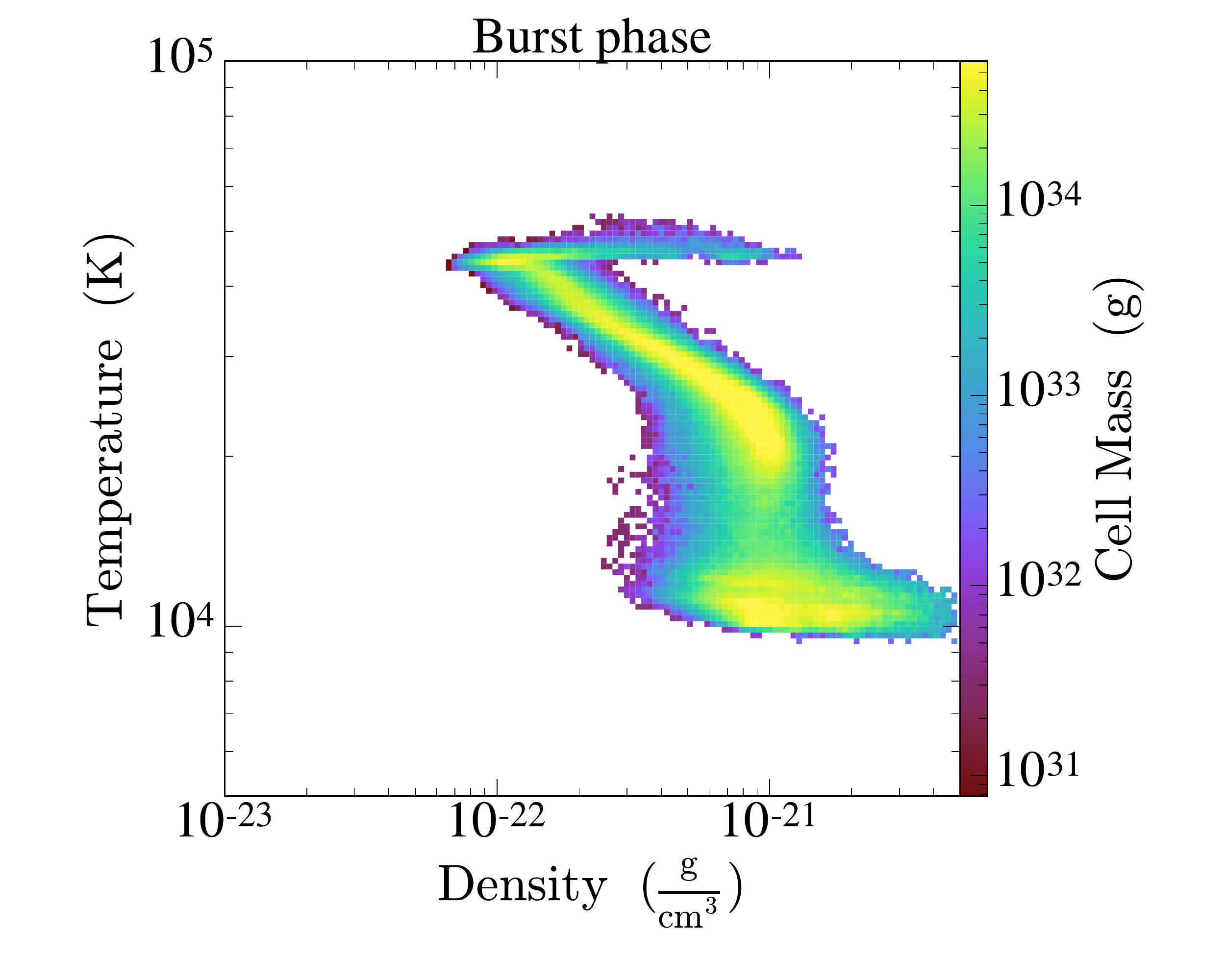}{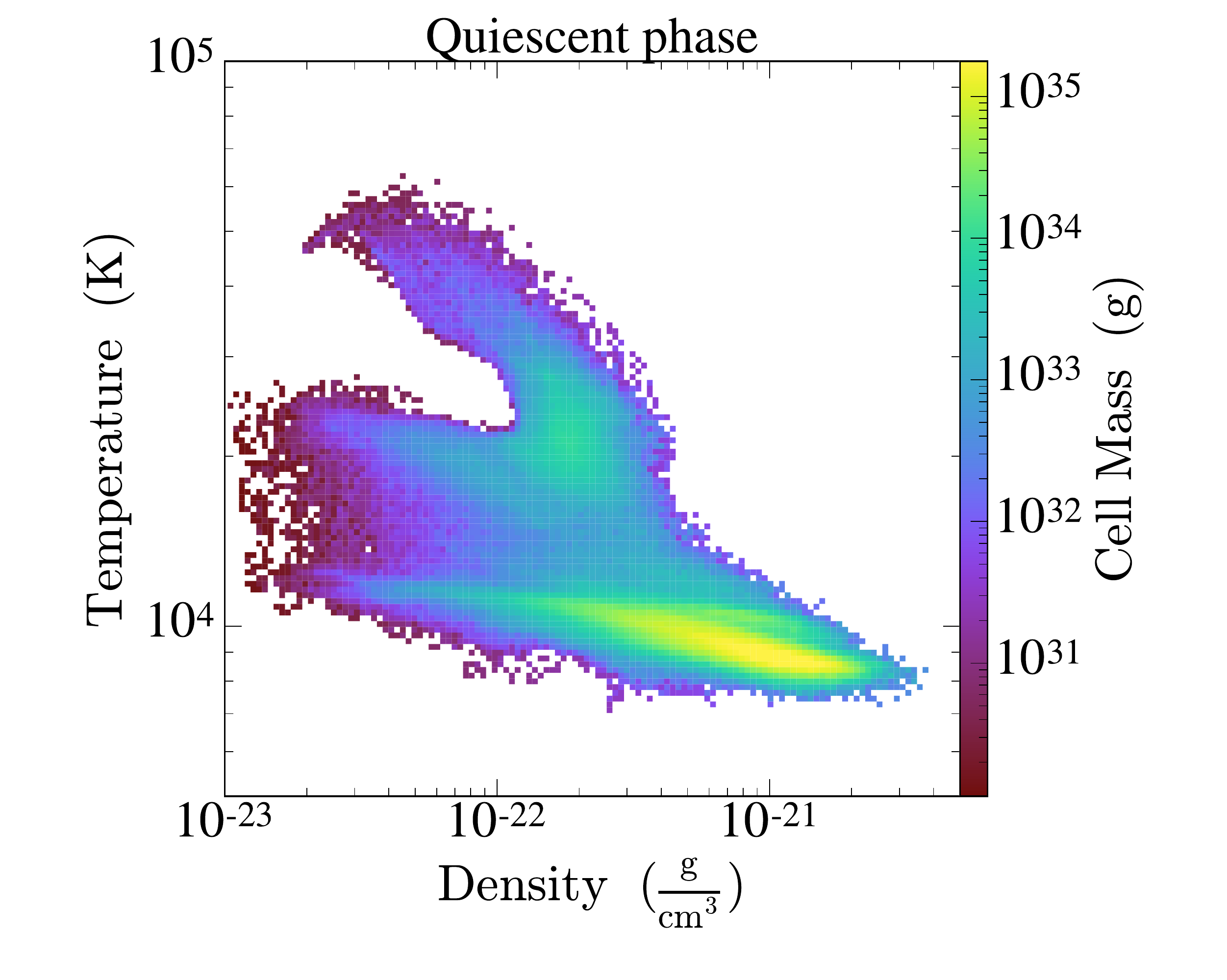}

\caption{Gas density and temperature distribution in \revI{M4N3E8mod}
run during the accretion burst (left) and quiescent (right) just
before the subsequent burst within the central \revI{8\,pc}. Color
marks the cumulative mass of the gas contributing to each cell of
the phase diagram.}

%Cell mass distribution as a function of gas density and
%  temperature of the simulation M6N1 at the burst (left) and just
%  before the burst (right) within a radius of 1\,kpc. During the
%  accretion bursts, most of the gas is photo-heated ($T \sim 4\times
%  10^4$\,K) near the BH and more effectively in the low density
%  gas that exists in the outer part of \str~sphere, where
%  temperature decreases with gas density. Just before the burst, low
%  temperature ($T \sim 10^4$\,K) gas dominates at all
%    densities. However, some gas still stays in the enhanced
%  temperature region ($T > 2\times 10^4$\,K). 
\label{fig:phase} \end{figure*}

% FIG1

Figure~\ref{fig:snapshot1} shows the evolution of the gas density (top),
\hii~fraction ($n_{\rm HII}/n_{\rm H}$, middle), and temperature (bottom) for
the \revI{M4N3E8mod} run. From left to right the snapshots illustrate evolution of the
ionized region starting with the burst of accretion and ending with a quiescent
phase just before the subsequent burst. In general, the region under the
influence of ionizing radiation is characterized by the low gas density, high
ionization fraction, and high temperature ($T \sim 4\times 10^4$\,K).

% Density

The low density \str~sphere roughly maintains spherical symmetry throughout a
sequence of oscillations despite a highly turbulent nature of the gas between
the high and low density regions separated by the ionization front.  In
contrast to the turbulent features imprinted in the density map shown in the
top panels of Figure~\ref{fig:snapshot1}, the \hii~fraction and temperature
maps are relatively uniform. All maps illustrate a correlation of the size of
the \str~sphere with the accretion rate: as the accretion rate decreases (from
left to right), the average size of \str~sphere also decreases. Unlike the
1D/2D simulations however, in 3D simulations the \str~sphere never completely
collapses between the accretion outbursts.

% HII 

The outline of the \hii~region traced by the \hii~fraction and temperature
shows a more dramatic departure from spherical symmetry than the smoother
density maps. The ``jagged" edge of the \hii~region can be directly attributed
to ionization of gas by the UV photons.  It arises as a consequence of
inhomogeneity of the gas driven by turbulence, which produces a range of
column densities along different radial directions, as seen by the central
source.  More energetic photons with longer mean free paths travel beyond the
ionization front of the \hii~region partially ionizing the gas there. Their
effect is noticeable in the \hii~fraction maps as a light green halo
surrounding the ionization front, \revI{size of which} remains approximately constant regardless of the accretion phase.

%% FIG2
\subsection{Accretion rate and period of oscillation}

The top panel of Figure~\ref{fig:acc_rate} shows evolution of the
accretion rate, which is calculated from the gas density (middle), and
gas temperature (bottom), for the runs M4N3 (blue solid), M4N3sec (green
dashed), and M4N3E8mod (red dashed). The temperature of the ionized
region, $T_{\rm HII}$, increases modestly during the accretion bursts
and consequently, does not play a large role in determining the
accretion rate onto the BH. It follows that the density of the ionized
region, which varies by a few orders of magnitude, is the primary
driver of evolution of the accretion rate.

Indeed, the amplitude and variability of the accretion rate in
Figure~\ref{fig:acc_rate} closely follow the density of
the ionized gas, as expected given the adopted prescription of the Bondi
accretion rate, $\dot{M}_{\rm BH} \propto \rho_{\rm HII}$. A notable
departure of the 3D simulations is that the accretion rate spans the
range of 2-3 orders of magnitude compared to the 1D/2D simulations,
where this range is measured to be 5-6 orders of magnitude
\citep{ParkR:11,ParkR:12}. The difference seen in the 3D simulations can
be directly attributed to a higher level of ``quiescent" accretion
that occurs between the outbursts, while the maximum and mean
accretion rate remain approximately unchanged relative to the 1D/2D
models. \revII{In the case of 1D/2D simulations, a sharp density
drop inside of the ionization front is well preserved until the
front collapses to a very small radius. The relatively high gas
density \revIII{inside the ionization front} in the 3D simulations,
explains the higher level of accretion rate} \revIII{during the
quiescent phase.} \revI{The difference between the current 3D and
former 1D/2D simulations may arise
due to different dimensionality of simulations, however we cannot
completely rule out a possibility that the difference in the codes
and numerical schemes are also partially responsible.}

We also investigate the effect of modified spectrum of the
ionizing source as well as that of secondary ionizations by more
energetic X-ray photons which generate energetic particles causing
further ionization. Specifically, in addition the baseline run
M4N3, we carry out the run M4N3sec, which accounts for secondary
ionizations and M4N3E8mod, which uses our modified
prescription for the ionizing continuum modeled in 8 energy bins. We
find consistent results for all three. Based on this we conclude that
(a) secondary ionizations do not have a significant impact on the
thermodynamic state of the gas and (b) that our prescription for the
ionizing continuum in M4N3E8mod returns physical results
indistinguishable from the baseline simulation (see Appendix for more
discussion of the latter).

In addition to the peak and mean accretion rates, the mean period
between the bursts of accretion (i.e., the accretion duty cycle) is
consistent with the previous 1D/2D results. The mean accretion rate
(dotted lines) for all M4N3 runs is approximately 2 orders of magnitude
lower than the Bondi rate \revI{for neutral gas} (dot-dashed line). 
The mean period between the
bursts is approximately $0.25\,$Myr, consistent with earlier results of \citet{ParkR:11, ParkR:12}
\begin{equation} \tau_{\rm cycle} \sim 0.22\,{\rm Myr}
\left(\frac{\mbh}{10^4\,\msun}\right)^{\frac{2}{3}}
\left(\frac{\nH}{10^3\,{\rm cm^{-3}}}\right)^{-\frac{1}{3}}
\label{eq:cycle}
\end{equation} 
with $\alpha=1.5$ and $\eta=0.1$ for radiative efficiency.

%% FIG4 %%%%%%%%%%%%%%%%%%%%%%%%%%%%%%%%%%%%%%%%%% 

\begin{figure*}[t]
\epsscale{1.2} \plotone{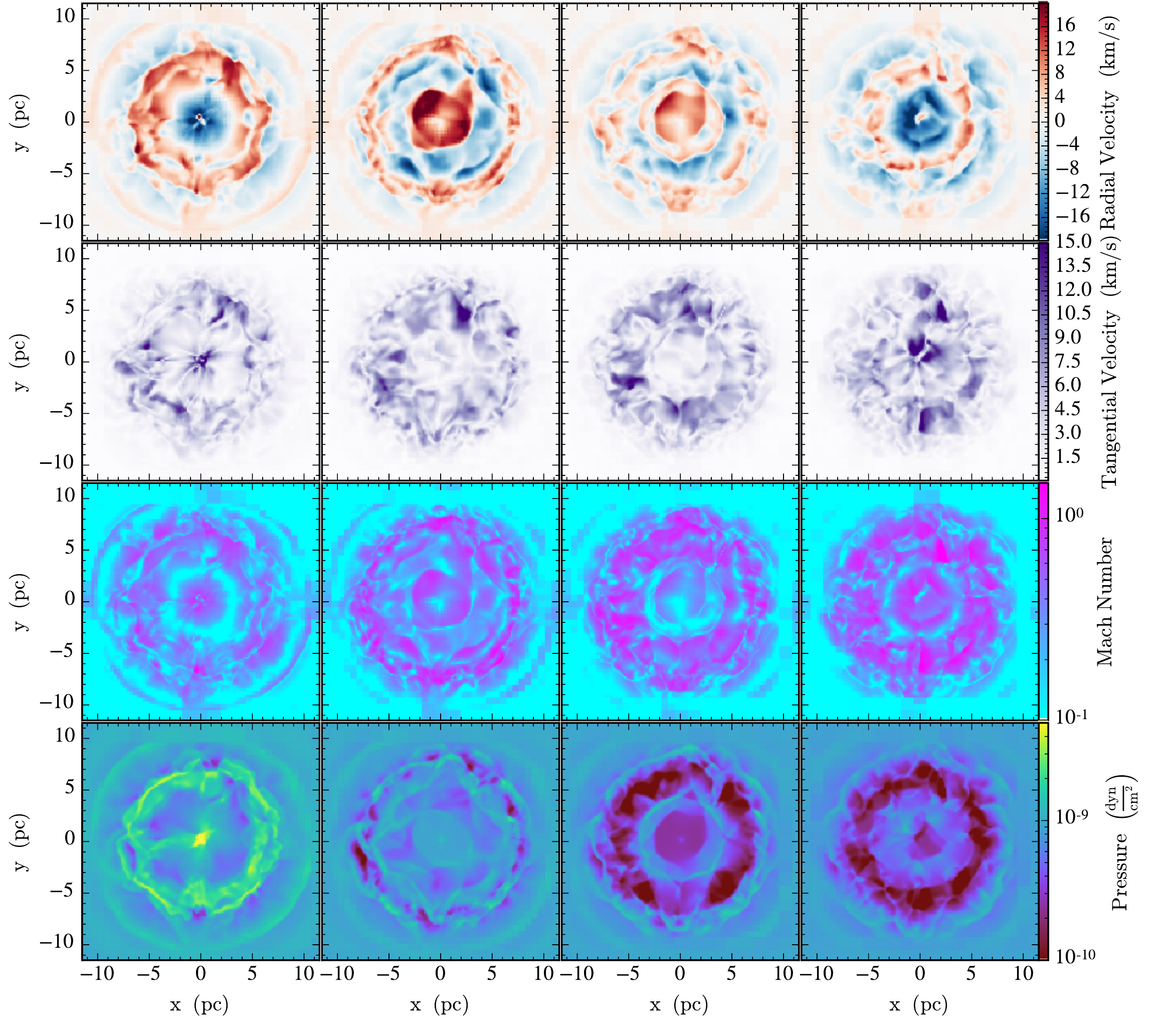}

\caption{From top to bottom: slices of radial velocity, tangential
velocity magnitude, Mach number, and thermal pressure in run
\revI{M4N3E8mod}, respectively. The times for each snapshot match
those shown in Figure~\ref{fig:snapshot1}. }

\label{fig:snapshot2} \end{figure*}

\subsection{Phase diagrams and velocity structure}
%%% FIG3

Figure~\ref{fig:phase} shows the phase diagrams of the gas density and
temperature during the burst (left) and quiescent (right) phases for 
\revI{M4N3E8mod} run.
The plotted values of density and temperature are measured within \revI{8\,pc}
radius surrounding the BH (see Figure~\ref{fig:snapshot1}).  During the
bursts, the gas in vicinity of the BH is photo-ionized and photo-heated to $T
\approx 5\times10^4\,$K, as illustrated by a horizontal branch in the $T-\rho$
distribution. The rest of the gas in the outer part of \str~sphere is shown
as a separate branch, characterized by temperature decreasing with density.
This temperature structure is consistent with Figure~\ref{fig:snapshot1} where
the central region shows a fairly uniform temperature, which then decreases
with radius at larger separations from the BH.

During the quiescent phase, most of the gas cools through recombination
to $T \approx 10^4\,$K. However, a smaller fraction of high
temperature, low density gas still appears in the same location at
the top left corner of the diagram, giving rise to a relatively
wide multi-phase distribution of gas. \revII{Note that in 1D/2D
simulations, the central region is replenished with high density
gas during the quiescent phase as the ionization front collapses
to the minimum size, which is significantly smaller
than the size of the \hii~region on average 
\citep{MiloCB:09, ParkR:11}.} In 3D
simulations, even though the ionization front does not collapse entirely during the quiescent phase \revII{(i.e., the radius of \hii~bubble remains approximately half
the average size)}, \revI{the enhanced
gas density} within the ionized region is still sufficient to trigger
the next burst of accretion. \revII{The reason why the hot bubble
maintains larger size during the quiescent phase in 3D simulations is
related to the higher accretion rate and enhanced gas density
just inside the ionization front, as explained in the previous section. }

%\revI{Note that the aforementioned Mode-II oscillation
%\citep{MiloCB:09, ParkR:12} also shows that the \hii~bubbles do not
%disappear even during the quiescent phase.} 

%%% FIG5 %%%%%%%%%%%%%%%%%%%%%%%%%%%%%%%%%%%%%%%%%%

\begin{figure*}[t] \epsscale{1.05}
\plotone{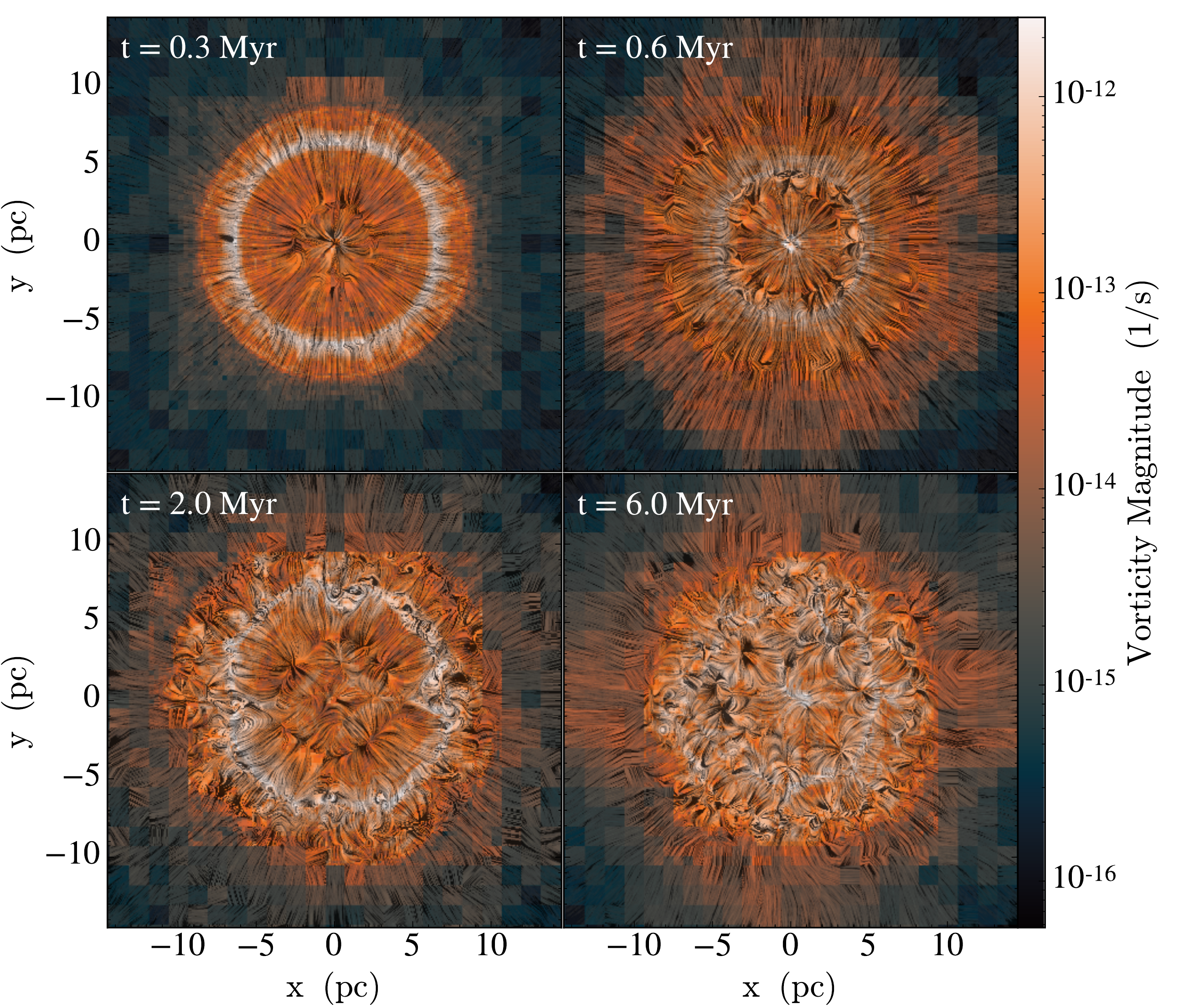}

\caption{Evolution of vorticity magnitude, $\magvolt$, in run
\revI{M4N3E8mod} at \revI{$t=$ 0.3, 0.6, 2.0, 6.0\,Myr} over-plotted
with the line integral convolution of velocity field
\revI{\citep{Cabral:1993} where the texture visualizes the local direction
of the velocity field}. The vorticity builds up over
time but saturates during the \revI{early
phase} of the simulation.}

\label{fig:vorticity} \end{figure*}

%%% FIG4

Figure~\ref{fig:snapshot2} shows the radial velocity, tangential
velocity magnitude, Mach number, and thermal pressure in run \revI{M4N3E8mod} during
the same phases shown in Figure~\ref{fig:snapshot1}. During the burst of
accretion (first column), the gas in the central region falls into the
BH boosting the accretion rate while a strong outflow (colored in red)
is still observed in the outer part of the ionized region at \revI{$r\sim
5$\,pc}. The outflow velocity reaches $\sim 20$\,km/s, the transonic
value for the temperature of the ionized region, as illustrated by the Mach number slices. 

As the transonic gas outflow encounters the
neutral medium, the kinetic energy of the gas is efficiently
dissipated by shocks. The supersonic gas is mostly found near the
ionization front or just outside of the \str~sphere. \revI{The
central core shows the highest thermal pressure during the burst
of accretion due to the high density of the core but the
temperature distribution within the ionized region remains uniform,
as shown in Figure~\ref{fig:snapshot1}}. A spherical shell around the ionization front
also exhibits high thermal pressure because the neutral clumps of
gas in this region are efficiently photo-heated by the ionizing UV
photons that escape absorption by the central core.

%The central core shows the highest
%thermal pressure as the dense gas close to the BH is photo-heated during
%the burst of accretion. 

After the burst, the central region starts to develop an outflow (second
and third columns) driven by the high thermal pressure. The high thermal
pressure, located at the core and around the ionization front in the
first column, expands and dissipates as the accretion powered
luminosity decreases in time. The average thermal pressure is maximal
during the burst (first column) and decreases as a function of time
displaying the minimum just before the subsequent burst (last column).
The inner region becomes gradually under-pressurized due to expansion
and radiative cooling. This allows the gas to flow toward the center,
causing a subsequent accretion burst.

The velocity structure in this work is distinct from the 1D/2D
simulations where a strong laminar (i.e., non-turbulent) outflow in the
outer part of the \str~sphere persists most of the time. In these
simulations the ionization front collapses completely due to the loss of
thermal pressure in the ionized region. The 3D simulations described
here instead show highly turbulent motion in both radial (first row) and
tangential directions (second row) that cascades to small scales over
several oscillation cycles. The  tangential velocity is only suppressed
in the central region during the strong inflow and outflow episodes.

%%%%% Evolution of vorticity
\subsection{Evolution of vorticity} 

%%% FIG5 

We use vorticity, defined as a curl of the velocity field
\begin{equation} 
\vec{\omega} = \vec{\nabla} \times \vec{v}
\label{eq:vorticity} 
\end{equation}
to quantify the degree of turbulent motion of the gas.
Figure~\ref{fig:vorticity} shows the evolution of vorticity at times
\revI{0.3, 0.6, 2.0, and 6.0\,Myr} in the run \revI{M4N3E8mod}. The
line integral convolution \revI{\citep{Cabral:1993}}, a method to
create a texture correlated in the direction of the vector field,
is shown in black over vorticity magnitude. The vorticity is highest
around the ionization front and propagates \revI{inward} and outward.
It builds up over time and saturates during the \revI{early phase}
of the simulation, at \revI{$t=2.0\,$Myr}, after which point it
does not display a significant increase until \revI{$t=6.0\,$Myr}.

The propensity of the gas to develop turbulence in 3D simulations
can be understood in the context of the vorticity equation, which
written as Lagrangian derivative ($D/Dt$) reads
\begin{equation}
\frac{D\vec{\omega}}{Dt} = \frac{\partial \vec{\omega}}{\partial t}
+(\vec{v} \cdot \nabla )\vec{\omega} = \frac{1}{\rho ^2}\nabla \rho
\times \nabla p .
\end{equation}

The right hand side of the equation quantifies the {\it baroclinicity}
of a stratified fluid, present when the gradient of pressure is
misaligned from the density gradient of the gas. In our
simulations, the local density gradients have no preferred direction
because of the turbulence which produces a significant inhomogeneity of
density, as evident in Figure~\ref{fig:snapshot1}. On the other hand,
the pressure gradient is generally along the direction of gravity and largest across the ionization front. This misalignment dictates the evolution of $\vec{\omega}$ close to the ionization front.

%\jhw{[Is this really true?  I would have
%thought that the pressure gradient is the largest going across the
%ionization front, given the temperature differences, and the density
%gradients have no preferred direction because of the turbulence.]}

From dimensional analysis the magnitude of vorticity squared can
be estimated as $|\vec{\omega}|^2 \sim G\mbh / r^3$, where the
relevant radius corresponds to the size of the \str~sphere.  
%
%\begin{equation}
%\left|\vec{\omega}\right|^2 = \left|\frac{v}{r}\right|^2 \sim \frac{G\mbh}{r^3}
%\end{equation}
%
%where $v=\sqrt{2G\mbh/r}$ corresponds to the \tb{free fall [check this]} speed. 
The number of ionizing photons from the BH is proportional to the
density squared from recombination rate as well as the recombination
volume, $N_{\rm ion} \propto \langle R_s \rangle^3 \nH^2$ where
$\langle R_s \rangle$ is the mean size of the \str~radius. On the
other hand $N_{\rm ion}$ is also proportional to the Bondi accretion
rate, i.e., $N_{\rm ion} \propto \mbh^2 \nH$ \citep{ParkR:11}. Thus,
the mean size of \str~sphere is related to the BH mass and gas
density as $\langle R_s \rangle^3 \nH^2 \propto \mbh^2 \nH$. Using
this in the estimate of the mean magnitude squared, $|\vec{\omega}|^2$,
which develops around the ionization front
\begin{equation}
\left|\vec{\omega}\right|^2 \sim \frac{G\mbh}{\langle R_s \rangle ^3}
\propto \frac{\nH}{\mbh}.  
\label{eq:w2}
\end{equation}

%%% FIG6 %%%%%%%%%%%%%%%%%%%%%%%%%%%%%%%%%%%%%%%%%%

\begin{figure}[t] \epsscale{1.2}
\plotone{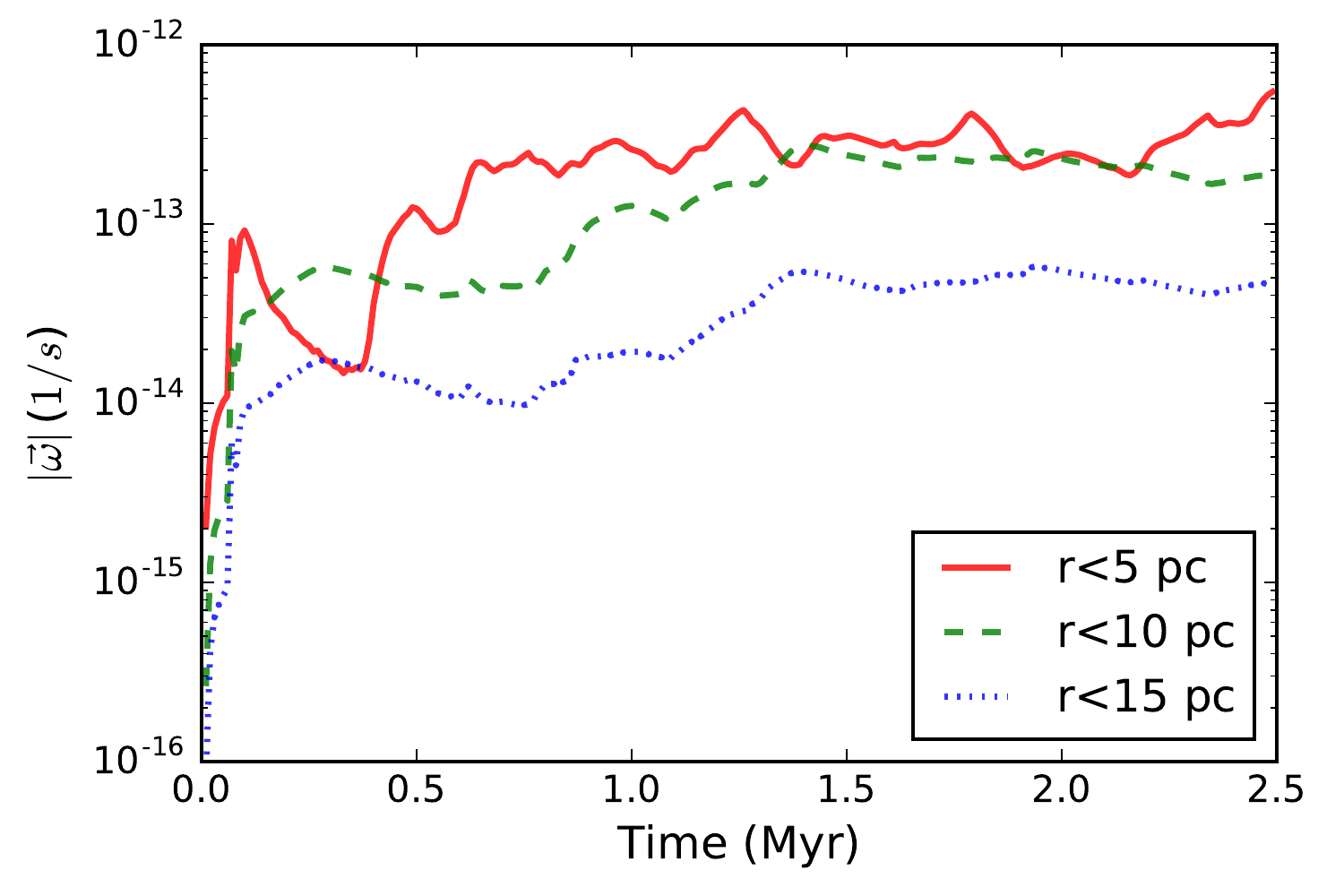}

\caption{Time evolution of the mass-weighted mean vorticity $\magvolt$ 
for run M4N3, calculated within the radius $r=5$, 10, and 15\,pc.}
\label{fig:w_time} \end{figure}

% FIG 6: 

Figure~\ref{fig:w_time} shows the time evolution of mass-weighted
mean vorticity $\magvolt$ for the simulation M4N3, calculated within
the radius $r=5.0$ (solid line), 10.0 (dashed), and 15.0\,pc (dotted).
\revI{On average, mean vorticity shows a rapid increase at the
beginning at all radii and remains at a constant level after $t
\approx 1.3$\,Myr. For the $r < 5.0$\,pc volume, the vorticity
shows a significant change during the initial phase ($t \lesssim 0.5$\,Myr)
, when the
ionization front crosses this scale, but reaches a steady value
with a minor variation after only $t \gtrsim 1$\,Myr. The small
variation in $\magvolt$ for the $r < 5.0$\,pc volume shows an approximate 
match with the accretion cycles
in Figure~\ref{fig:acc_rate}. On the other hand, $\magvolt$ for the 
volumes with $r < 10.0$\,pc and $< 15.0$\,pc show a steady increase until
$t \approx 1.3$\,Myr and remains at the same level after that. Note
that for the run M4N3 the ionization front extends to a radius of
$r\approx 10$\,pc as shown Figure~\ref{fig:snapshot1}, thus $r\approx
10$\,pc volume is large enough to capture most of the vorticity. 
The value of $\magvolt$ for $r < 5.0$\,pc and $r < 10.0$\,pc volumes after
$\gtrsim 1.3$\,Myr becomes comparable, which means that there is
no noticeable difference in $\magvolt$ in inner and outer parts of the ionized
region. However, as we increase the volume, i.e., $r<15.0$\,pc,
the mean $\magvolt$ decreases since the volume outside
of ionized region with low $\magvolt$ is included. }

\begin{figure}[t]
\epsscale{1.20}
\plotone{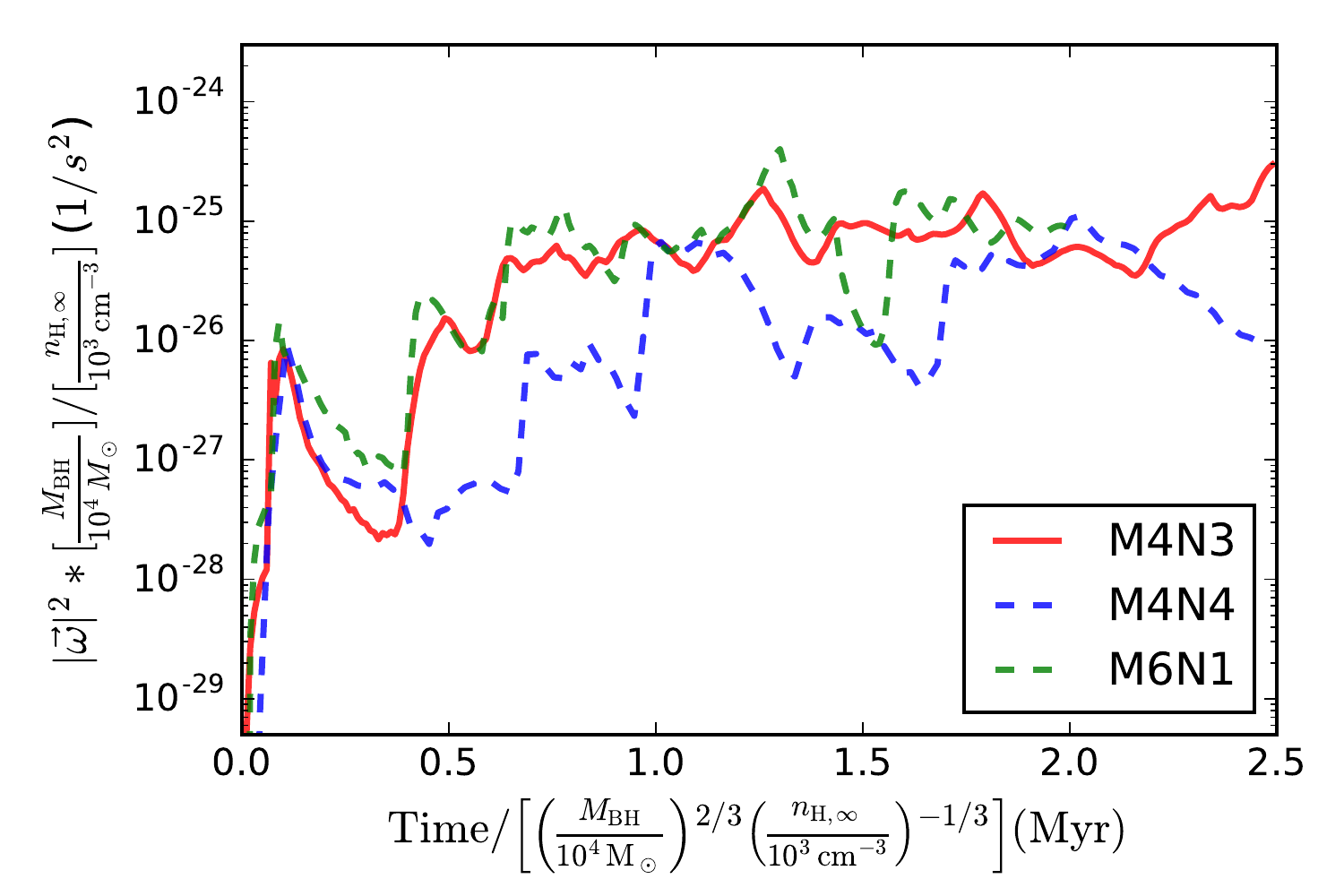}

\caption{Evolution of the mass-weighted mean vorticity squared for runs
M4N3 (solid red), M4N4 (dashed blue), and M6N1 (dashed green) calculated
within the radius of $0.5 \langle R_s \rangle$. \revI{$\magvolt^2$} is
normalized by $\nH/\mbh $ and time is in units of $(\mbh^2/\nH)^{1/3}$
(see text for more details).} 

\label{fig:w2_den} \end{figure}

Figure~\ref{fig:w2_den} shows the evolution of the mean vorticity
squared for simulations M4N3 (solid red, 5\,pc), M4N4 (dashed blue,
2.5\,pc), and M6N1 (dashed green, 0.5\,kpc) within the radius written in
parentheses for each run. The selected radius is approximately a half of
the mean \str~radius $\langle R_s \rangle$ for each. To show all three
simulations on the same scale, we normalize
$\left|\vec{\omega}\right|^2$ by $\nH \mbh^{-1}$, as in
Equation~(\ref{eq:w2}), and the time on the $x$-axis by the average
length of the oscillation cycle $\propto\mbh^{2/3} \nH^{-1/3}$
shown \revI{in Equation~(\ref{eq:cycle})}. 
%\tb{[I would skip the following statement, they
%all seem pretty much consistent.]} 
%\khp{[I just want to be more careful
%by keeping the sentence since M4N4 is not a family of M6N1 and M4N3.]
Note that simulations M4N3 and M6N1 which share the same value of
$\mbh\nH=10^7\,\msun {\rm cm}^{-3}$, display a good match while \revI{M4N4}
shows a reasonably consistent result considering the large range of
$\left|\vec{\omega}\right|^2$.

%%% FIG7 %%%

% FIG8

\subsection{Turbulent Kinetic Energy} 

In order to quantify how much turbulent energy is present on different
spatial scales we calculate the energy spectrum $E(k)$ as a function of
the wavenumber $k$. The total turbulent kinetic energy per unit mass is
evaluated as
\begin{equation}
\frac{1}{2} \left< |\vec{v}|^2 \right> = \int_{0}^{\infty} E(k)dk 
\label{eq:tke}
\end{equation}
where $\vec{v}$ shown here only accounts for the turbulent component
of the velocity.

%and thus the turbulence is homogeneous and isotropic. 

\begin{figure}[t] \epsscale{1.23}
\plotone{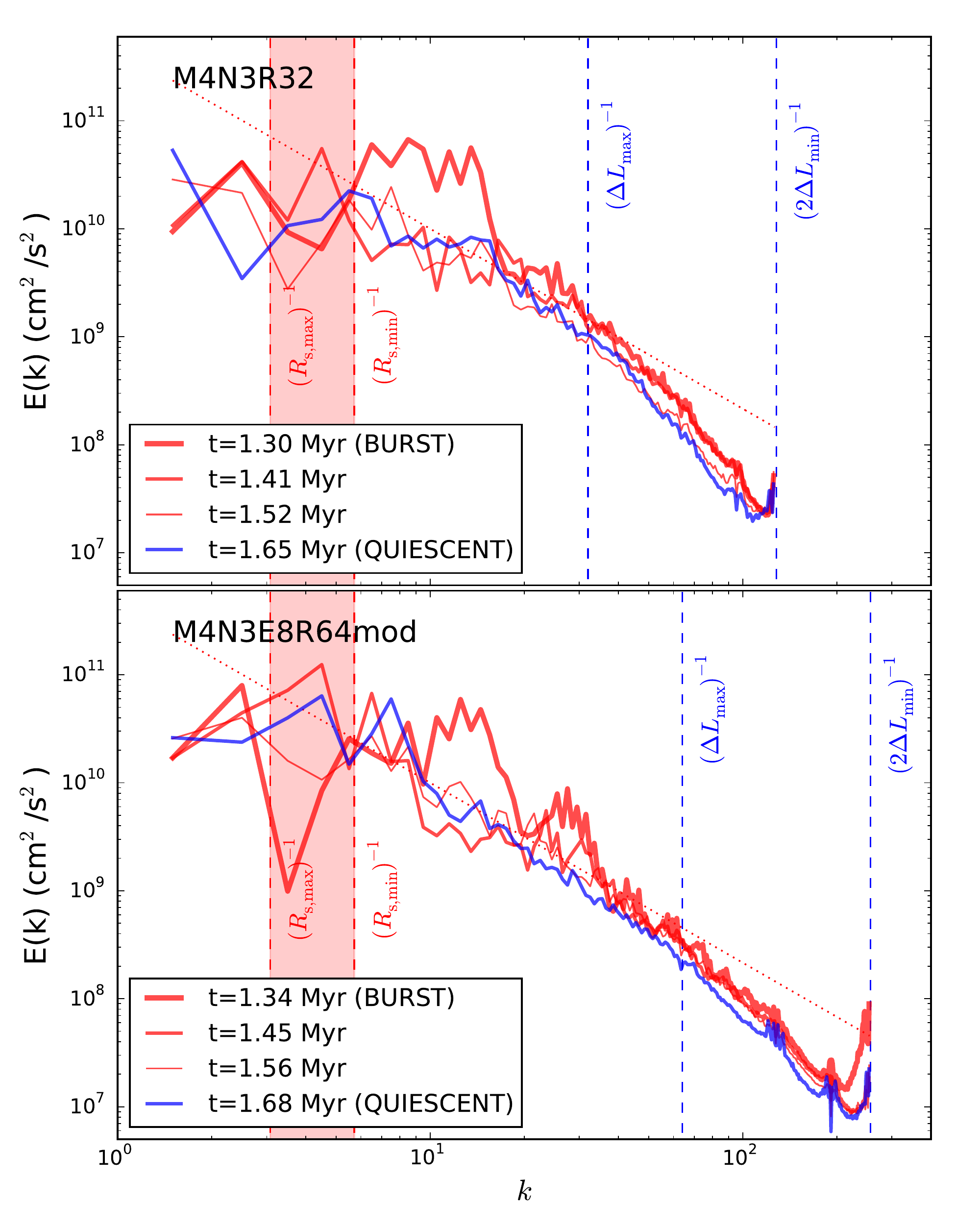}

\caption{Power spectrum of the specific turbulent kinetic energy at
different times during one accretion cycle in simulation 
\revI{M4N3R32 (top) and M4N3E8R64 (bottom)}. 
$R_{\rm s, max}$ and $R_{\rm s, min}$ are the maximum and minimum
radius of the \str~sphere and $\Delta L_{\rm max}$ and $\Delta L_{\rm
min}$ are the sizes of the coarsest and the finest resolution element, respectively. The dotted line traces the Kolmogorov spectrum with the slope $E(k) \propto k^{-5/3}$. }  
\label{fig:powerspec} 
\end{figure}

Figure~\ref{fig:powerspec} shows the specific turbulent kinetic
energy power spectrum from runs \revI{M4N3R32 (top) and M4N3E8R64mod
(bottom)}, plotted for phases of an accretion cycle \revI{at $t\sim
1.3$\,Myr after which the mean vorticity reaches at a steady value
as shown in Figure~\ref{fig:w_time}}. The wavenumber is defined as
$k = 1/\Delta L$, where $\Delta L$ corresponds to the scale
ranging from the finest resolution element to the size of the entire
computational domain. In calculations presented here $\Delta L$
(and hence, $k$) is expressed in code units while $E(k)$ is shown
in ${\rm cm}^2/{\rm s}^2$.  The shaded region defined by $R_{\rm s,
max}$ and $R_{\rm s, min}$, which represent the maximum and minimum
sizes of the \str~radius, marks the scale for the turbulence
source. $\Delta L_{\rm max}$ and $\Delta L_{\rm min}$ indicate the
resolutions of the top and finest grids, respectively. 
\revI{Note that the high resolution run M4N3E8R64mod is characterized by a 
wider range of k which extends to larger values, resulting in a better 
resolved turbulence at small spatial scales.}

%%% FIG8 %%%

%%% FIG9 %%%

\begin{figure}[t] \epsscale{1.2}
\plotone{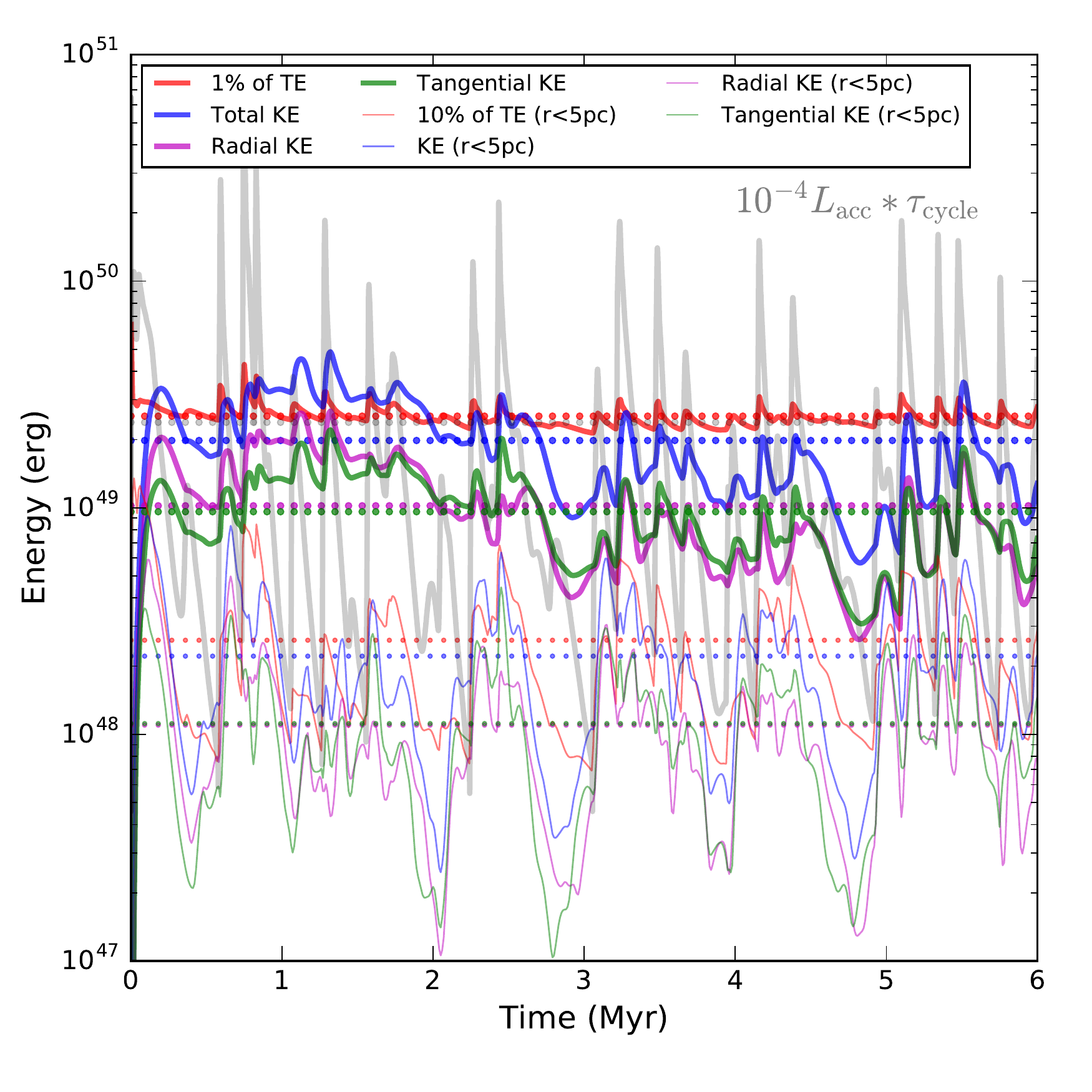}

\caption{Evolution of the total thermal (red) and kinetic (blue)
energy for M4N3E8. Gray line shows the energy in radiation emitted
from the BH during one cycle of oscillation (i.e., $L_{\rm
acc}*\tau_{\rm cycle}$). Kinetic energy is decomposed into radial
(magenta) and tangential (green) components. Horizontal dotted
lines indicate mean values for each energy component. \revI{Thick and
thin lines show the energies in the entire volume and within the
sphere with radius $r=5$\,pc, respectively.}}

\label{fig:energy_evol} \end{figure}

The turbulence is sourced on the scales that are just inside the
ionization front ($k \la 6$) and it cascades to smaller scales
(larger $k$). Figure~\ref{fig:powerspec} illustrates that in
quiescence (blue line) the turbulent energy tends to be \revI{slightly}
lower across the spectrum relative to the other phases of the
accretion cycle. The turbulent energy spectrum calculated from
simulations is broadly consistent with the Kolmogorov spectrum
[$E(k) \propto k^{-5/3}$; \citet{Kolmogorov:91}], which provides
analytic description of a saturated and isotropic turbulence. 
Departure from the Kolmogorov spectrum is noticeable in two spatial
regions in Figure~\ref{fig:powerspec}. Namely, the turbulence is
diminished in neutral gas, outside of the ionization sphere, which
is consistent with our hypothesis of radiatively driven turbulence.
\revI{Secondly, the simulated spectrum departs from Kolmogorov at wave
numbers $k \gtrsim 30$ for M4N3R32 and $k \gtrsim 60$ for
M4N3E8R64mod, respectively, that correspond to scales smaller than
\revI{$\Delta L_{\rm max}$}.} \revI{Note that the inertial regime
where $E(k) \propto k^{-5/3}$ is extended to larger $k$ for
M4N3E8R64mod whereas the dissipation regime arises at $k \gtrsim
\Delta L_{\rm max}^{-1}$ consistently for both simulations with different
resolutions.} This region coincides with a portion of a simulation
domain where we employ several different refinement levels in the
form of layered grids. Specifically, while the highest level grid
resolves eddies of the size
$2\Delta L_{\rm min}$, the base grid will resolve eddies with the
minimum size of $\Delta L_{\rm max} = 8\Delta L_{\rm min}$.
Because of the non-uniform sampling of turbulence in these regions some
fraction of the turbulent kinetic energy is numerically dissipated
before it cascades to the smallest scales. This expectation is
consistent with the spectrum that steepens towards smallest scales
(highest $k$ numbers), as shown in the figure. 
The turbulent kinetic
energy contained on small scales is however a small fraction of the
total turbulent kinetic energy and we do not expect that numerical
dissipation of turbulence into the internal energy of the gas will
significantly affect gas thermodynamics on these scales.

% \khp{Interestingly, this dissipation range coincides with the scale smaller than the top grids $\Delta L_{\rm max}$ meaning that we can not exclude the chance that the AMR scheme might not fully resolve the turbulence. However, the turbulent kinetic energy on small scales is a small fraction of the total turbulent kinetic energy, as discussed next.}

%so, the turbulence will dissipate from shocking. In any case,
%numerical dissipation will have an effect near the resolution
%limit.)} \khp{(When the $n_\rho = 1/3$ to calculate $E(k)$ for
%stratified medium, $E(k)$ in low k touches the transonic values, but
%the current figure with $n_\rho = 0$ shows much lower values in low
%$k$. However, $n_\rho = 0$ shows a great match between TKE and KE.)}

% FIG9

%\tb{[This section should be the punchline of the results but somehow,
%they do not emerge clearly and seem lost in the details of multiple
%lines and energy components. Lets try to sharpen the argument and
%emphasize the important stuff. Here is my take, feel free to modify.]}

Figure~\ref{fig:energy_evol} summarizes contribution to the total energy
budget of the simulated gas flow from the accretion powered radiation,
thermal and kinetic energy of the gas in run M4N3E8. We find that the
average energy of radiation in one accretion cycle, $\langle
L_{\rm acc}\,\tau_{\rm cycle}\rangle \sim 2.4\times 10^{53}\,$erg,
dominates over all other components of energy by more than
$\sim$\,2 orders of magnitude. The energy of radiation is estimated
as the bolometric luminosity emitted from the BH multiplied by a
characteristic length of one cycle of oscillation and is plotted as a gray
line. This implies that radiation can easily drive turbulence and
account for the increased internal energy of the gas, as energetic
requirements for these are comparatively modest.

The remaining components of energy are the thermal energy of the gas
(TE), marked as red line in Figure~\ref{fig:energy_evol}, and the
kinetic energy (KE), marked as blue line. Because the relative
contributions to thermal and kinetic energy are different for the
strongly irradiated gas in vicinity of the BH and neutral gas outside of
the \str~sphere, we show the evolution of TE and KE in the entire
computational domain as well as within the sphere with radius $r=5$\,pc.
Recall that in the run M4N3E8 the ionization
front resides at a radius of $r\approx 10$\,pc, and so the volume
within $r<5$\,pc traces plasma enclosed within the \str~sphere at all
times.

The kinetic energy is a combination of the kinetic energy of the bulk
flow (i.e., the radial inflow and outflow of the gas) combined with
random motions of the gas due to turbulence. The total kinetic energy is
calculated from simulations as
\begin{equation}
{\rm KE} = \frac{1}{2} \Sigma m_i (v_{i,x}^2+v_{i,y}^2+v_{i,z}^2)  
\label{eq:ke}
\end{equation}
where $m_i$ is the cell mass and $v_{i,x}$, $v_{i,y}$, and $v_{i,z}$
are the velocity components in x, y, and z directions, respectively.
The average energies (horizontal dotted lines) show that the kinetic
energy (\revI{thick} blue) is equivalent to $\sim1\,\%$, of the thermal energy
(thick red) in the entire computation domain. Similarly, kinetic energy corresponds to $\sim7\,\%$ of the thermal energy for the gas
within $r<10$\,pc and to $\sim9\,\%$ for the gas within $r<5$\,pc
(the latter is shown with thin lines). Therefore, the kinetic energy of the gas corresponds to less than 10\% of the thermal energy anywhere in the computational domain. It then follows that the kinetic energy due to turbulence
cannot be the main contributor to the pressure support of the
gas, which is mostly provided by thermal pressure. Along similar lines,
even if all kinetic energy of the gas is promptly thermalized, it would
not significantly alter the internal energy of the gas. This is the
basis for our earlier statement that numerical dissipation of turbulence
due to finite spatial resolution should not affect thermodynamics of the
gas.

In order to estimate the contribution to the total kinetic energy from
the bulk flow and turbulence we initially decompose the kinetic energy
into the radial and tangential components, shown as magenta and green
lines in Figure~\ref{fig:energy_evol}, respectively. Specifically, we
calculate the radial component as KE$_{\rm R} = \Sigma m_i v_{i,r}^2/2$,
where $v_{i,r}$ is the radial velocity, and the tangential component as
KE$_{\rm T}$= KE - KE$_{\rm R}$. Figure~\ref{fig:energy_evol} shows that
shortly after the beginning of the simulation the two components achieve
equipartition and KE$_{\rm R}$ $\approx$ KE$_{\rm T}$. Because the bulk
flow of the gas is mostly along the radial direction, KE$_{\rm R}$ actually accounts for both the kinetic energy of the bulk flow and turbulence.
The KE$_{\rm T}$ component of kinetic energy is on the other hand mostly
contributed by turbulence. We estimate the magnitude of the turbulent
component of kinetic energy in radial direction as $\sim$ 1/2\,KE$_{\rm
T}$, given that turbulence is isotropic, so that tangential and radial
motions account for two and one degrees of freedom, respectively. It
follows that the total kinetic energy is KE = KE$_{\rm bulk}$ +
3KE$_{\rm T}$/2 and because KE$_{\rm R}$ $\approx$ KE$_{\rm T}$, the
turbulent kinetic energy contributes $\sim$\,3/4 of the total kinetic
energy. Therefore, turbulent motions on all scales dominate the kinetic
energy of the entire accretion flow.

% SECTION : discussion %

\section{Discussion and Conclusions}
\label{sec:discussion}

The main aim of this paper is to extend numerical studies of accretion
mediated by radiative feedback to full 3D local simulations and identify
physical processes that have not been captured by the local 1D/2D
models. We achieve this by performing 3D simulations of
radiation-regulated accretion onto massive BHs using the AMR code {\it
Enzo} equipped with the adaptive ray tracing module {\it Moray}
\citep{Wise:2011}. Our main findings are listed below.

\begin{itemize}

\item Our 3D simulations corroborate the role of ionizing radiation in
regulation of accretion onto the BHs and in driving oscillations in the
accretion rate. We also confirm the mean accretion rate and the mean
period between accretion cycles found in earlier studies
\citep{ParkR:11,ParkR:12}. Since this study adopts different code and
numerical schemes relative to the earlier ones, this provides
verification that 3D simulations described here accurately reproduce the
key features of 1D/2D models.

% show a similar cycle of accretion and feedback found in 1D/2D simulations.  Radiation emitted by a BH couples to the neighboring gas and increases its thermal energy by photo-heating and photo-ionization. The low density \hii~region formed by the accretion powered ionizing UV and X-ray photons regulates the gas supply causing the accretion rate to exhibit oscillatory behavior, due to a loop between accretion and feedback. Moreover, current work shows that the fluctuation of the \hii~region due to changing accretion luminosity is effective in seeding transonic turbulence around the ionization front.

%\item Our 3D simulations produce results consistent with the previous 1D/2D simulations \citep{ParkR:11,ParkR:12}. The mean accretion rate (i.e., a few percent of $\dot{M}_{\rm B}$) and the mean period between accretion cycles $\tau_{\rm cycle}$ approximately match the values presented in \citet{ParkR:11}.

\item The 3D
simulations show significantly higher level of accretion during the
quiescent phase \revIII{due to the enhanced gas density within the ionization 
front and thus} oscillations in the accretion rate of only $\sim
2-3$ orders of magnitude in amplitude, significantly lower than in
1D/2D models. \revI{We caution however that even though 3D simulations
seem to faithfully reproduce the 1D/2D simulations, we cannot fully
rule out a possibility that higher level of accretion during
quiescence is caused by differences in the numerical scheme employed
in this work.}

%As a consequence, the 3D simulations show 
%significantly higher level of accretion between the accretion outbursts 
%and oscillations in the accretion rate of only $\sim 2-3$ orders of
%magnitude, significantly lower than in 1D/2D models.

\item In terms of the energy budget of the gas flow, we find that the
radiative energy dominates over the thermal and kinetic energy of the
gas by more than {$\sim$\,2} orders of magnitude, implying that
radiation can easily drive turbulence and account for the increased
internal energy of the gas. The thermal energy of the gas dominates over
the kinetic energy by a factor of $\sim 10-100$ depending on the distance from the BH, while the kinetic
energy itself is mostly contributed by the turbulent motions of the gas.
The turbulence therefore does not contribute significantly to the
pressure support of the gas.

\end{itemize}

%\khp{This work solves radiative transfer equations in a multi-frequency
%regime which covers from $13.6$\eV to $100$\,keV. We use $N_\nu=4$
%and 8 which and we find that they show consistent results. }

%We limit our understanding of the current study only to the {\it
%feedback-dominated} regime where the Bondi radius is much smaller than
%the ionization front holds \citep{PacucciVF:2015}. As we explore the
%higher values of $\mbh \nH$, the accretion flow makes a transition to
%hyper-accretion regime in which the role of BH radiation plays only a
%minor role in regulating gas accretion
%\citep{Begelman:79,ParkRDR:14a,Inayoshi:2016,Park:2016}. It might be a
%natural consequence that the turbulence driven by the fluctuating
%\hii~region does not develop significantly for hyper-accretion flow
%since there is no driving source. \tb{[This paragraph seems like a very
%detailed statement for the concluding remarks. Besides, we haven't
%mentioned the hyper-accretion regime anywhere, so it is not obvious that
%we should do that here. I would omit it. Instead, we may conclude by
%providing an outlook on the area that studies BH seed growth in this
%paragraph, and by connecting to an even broader subject - cosmology - in
%the next paragraph.]} 

The local simulations of radiation feedback mediated accretion onto BHs
presented here are the first step towards investigation of the full 3D
phenomena that take place in vicinity of growing BH.  Future, more
sophisticated local 3D simulations will in addition to turbulence need
to capture the physics of magnetic fields, angular momentum of gas, and
anisotropy of emitted radiation. All these ingredients are likely to
play an important role in determining how quickly BHs grow and how they
interact with their ambient mediums.

This study also bridges the gap between the local and cosmological
simulations. In large cosmological simulations radiative feedback from
BHs is often treated as purely thermal feedback and calculated without
directly solving the radiative transfer equations. This is a practical
compromise because direct calculations of radiative transfer are still
relatively computationally expensive, albeit not impossible. This study
paves the way for the next generation of cosmological simulations in
which the BH radiative feedback will be evaluated directly. It remains
to be explored how the rich phenomena discovered in local simulations
will affect the details of the growth and feedback from BHs at the
centers of galaxies or BHs recoiling during galaxy mergers
\citep[e.g.,][]{Sijacki:11, SouzaLima:2017, Park:2017}.

%\citep[i.e.,][]{DiMatteo:08,DiMatteo:12} \tb{[I would skip the
%reference, so it does not seem that you are calling Tiziana out. Also,
%this is true more generally.]} 
%\jhw{(I would like to see a final paragraph that gives an outlook of
%  this problem and future applications from these results, i.e. what
%  needs to be done to incorporate our results into models and which
%  physical scenarios would benefit from better modelling.}

\acknowledgements 

This work is supported by the National Science Foundation (NSF)
under the Theoretical and Computational Astrophysics Network (TCAN)
grants AST-1332858, AST-1333360, AST-1333514. \revI{The authors
thank the anonymous referee for thoughtful and
constructive comments and suggestions. 
The authors also thank Jarrett Johnson for carefully reading our manuscript.}
KP and TB are in part
supported by the National Aeronautics and Space Administration
through Chandra Award Number TM7-18008X issued by the Chandra X-ray
Observatory Center,
which is operated by the Smithsonian Astrophysical Observatory for and
on behalf of the National Aeronautics Space Administration under
contract NAS8-03060. JHW is also supported by NSF grant AST-1614333, NASA grant NNX17AG23G, and
Hubble Theory grants HST-AR-13895 and HST-AR-14326. Support for programs
\#13895 and \#14326 were provided by NASA through a grant from the Space
Telescope Science Institute, which is operated by the Association of
Universities for Research in Astronomy, Inc., under NASA contract NAS
5-26555.  This work was performed using the open-source {\sc Enzo} and
{\sc yt} \citep{Turk:2011} codes, which are the products of
collaborative efforts of many independent scientists from institutions
around the world. Their commitment to open science has helped make this
work possible.

%%% FIG10 %%%

\begin{figure}[t] \epsscale{1.25} 
\plotone{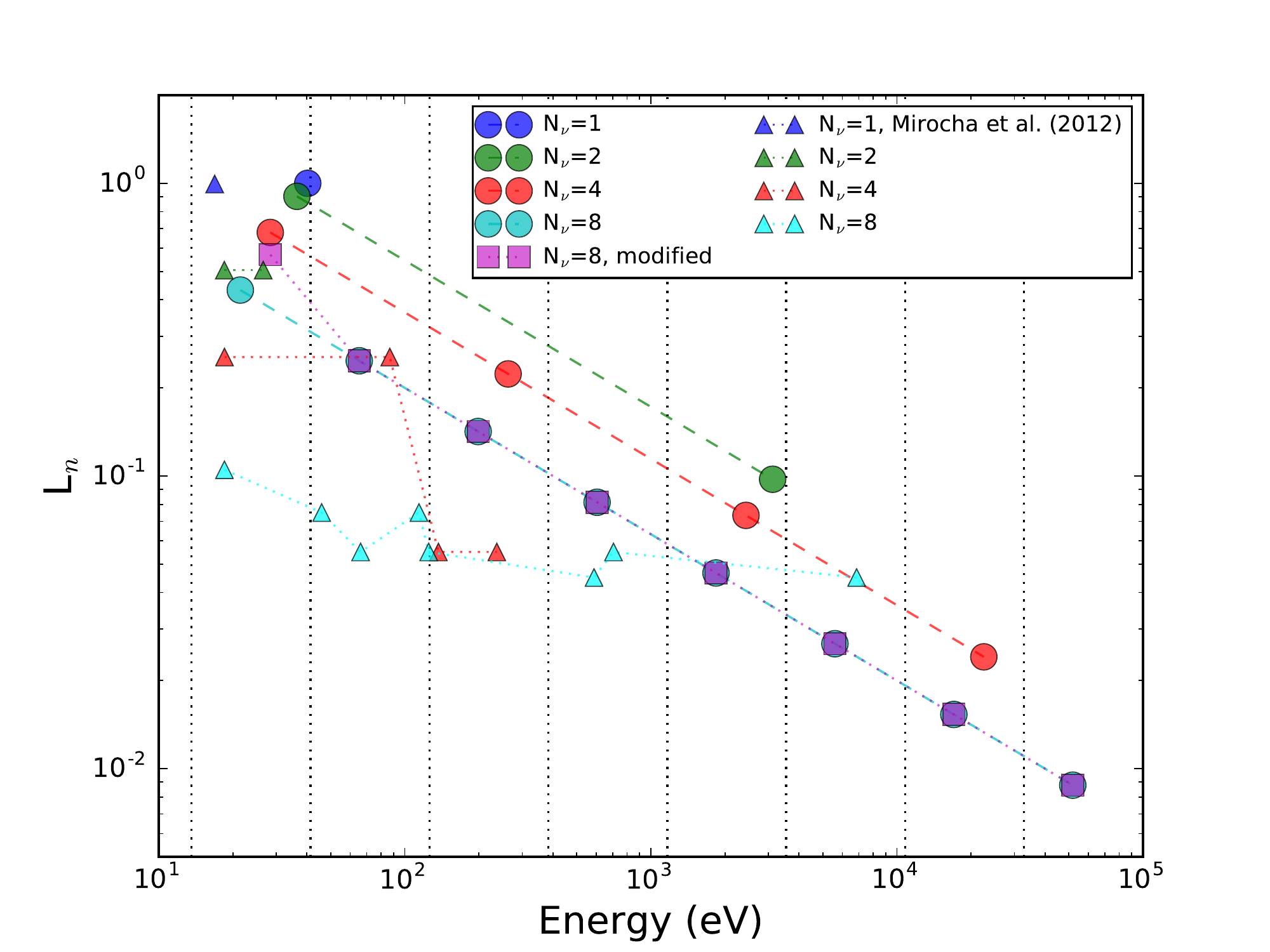}
\caption{Spectral energy distribution for a power-law spectrum with
$\alpha=1.5$ with different number of energy bins $N_\nu=1,2,4$, and
8. To match the $N_\nu=8$ case to $N_\nu=4$,
the first energy bin is shifted from 21.5\,eV to 28.4\,eV which
matches the energy for the case $N_\nu=4$. We also modify the
fraction of allocated energy for this bin from 0.4318 to 0.5704 so
that the number of total ionizing photons remains constant. \revI{Triangles show
the optimized SEDs using the method by \citet{Mirocha:2012}.}} 
\label{fig:sed}
\end{figure}

\appendix
\label{sec:appendix}
\section*{Modeling of the Power-law Spectral Energy Distribution}

We find that simulations, which rely on approximate description of the spectral energy distribution (SED) with a finite number of energy bins, are very sensitive to the precise configuration of those energy bins. Specifically, slightly different energy bin combinations result in a different temperature inside the \hii~region, which in turn affects the accretion rate. 

In this study we use an SED defined with $N_\nu = 4-8$ energy bins. Here, we elucidate how the energy groups are selected across the broad energy spectrum from 13.6\,eV to 100\,keV. Note that the mean energy of the entire photon distribution for power-law spectrum with $\alpha=1.5$ is $\langle E \rangle \sim 40.3$\,eV (blue symbol
in Figure~\ref{fig:sed}). Therefore, in SEDs modeled with $N_\nu \le 8$ energy bins, the lowest energy bin alone
covers the energy range $E \le \langle E \rangle$ (see Figure~\ref{fig:sed}). A majority of ionizing UV 
photons is assigned to the lowest energy bin, which in simulations are  trapped within the ionized low density \hii~region. High energy photons, which are characterized by longer mean free paths, travel beyond the ionization front partially ionizing the neutral gas there, as shown in the ionization fraction map of Figure~\ref{fig:snapshot1}. The temperature profile {\it within} the \hii~region is therefore sensitive to the number and energy of the UV photons, encoded in the energy of the first SED bin. 

On the other hand, the mean size of the \str~sphere, and thus the mean period of oscillation, is determined by the total number of ionizing photons \citep[see][for details]{ParkR:11}. Therefore, (1) the choice of the energy of the first SED bin and (2) the total number of ionizing photos must be considered together to produce consistent results from SED models with different spectral energy bin configurations.

%%% FIG 11 %%%

\begin{figure}[t] \epsscale{1.2}
\plotone{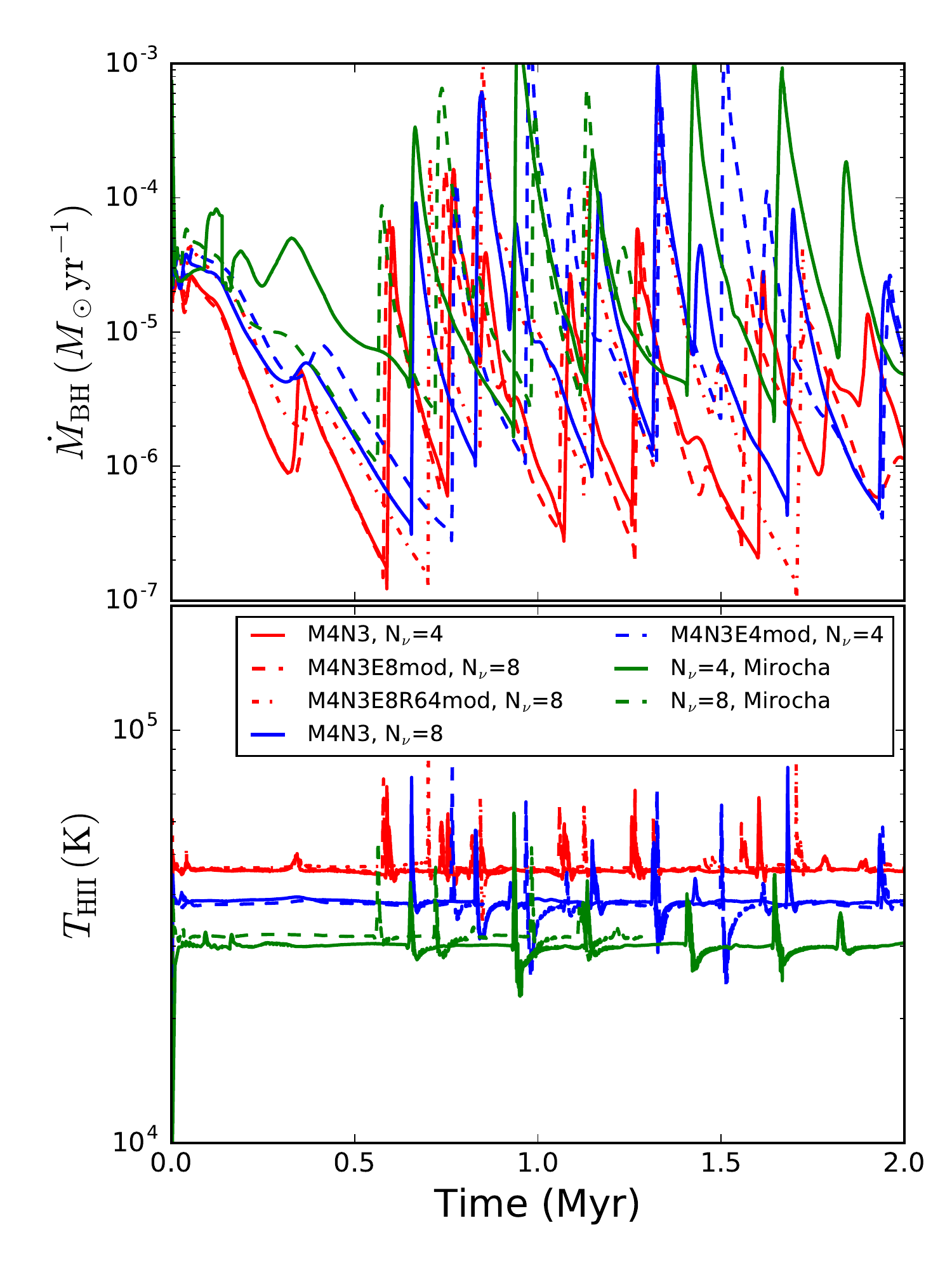} 
\caption{Accretion rate of BH (top panel) and temperature (bottom
  panel) of the \hii~region for different
  configurations. The same color indicates the similar results
  from different setups. For example, red solid line shows the
  model with $N_\nu=4$ with the equally spaced energy bins while
  red dashed line shows the model with $N_\nu=8$ with the
  modified SED. Both show a similar temperature and accretion
  rate evolution. The green lines show the results for SED
  calculated using the method in \citet{Mirocha:2012} for $N_\nu=4$
  (solid) and 8 (dashed).}
\label{fig:sed_test}
\end{figure}

%%% TABLE A %%%%%%%%%%%%%%%%%%%%%%%%%%
\begin{table}[thb]
\begin{center}
\caption{Optimized SEDs for $\alpha=1.5$ Power-law BH radiation using \citet{Mirocha:2012}}
\begin{tabular}{ccccc}
\hline 
\hline
$n_\nu$    & $N_\nu=1$	& $N_\nu=2$	    & $N_\nu=4$	& $N_\nu=8$\\	
\hline
1 & (16.88,\,0.995) & (18.47, 0.505) & (18.50, 0.255) & (18.49, 0.105) \\
2 & ...    	      & (26.58, 0.505) & (86.90, 0.255) & (45.94, 0.075) \\ 	
3 & ... 	      &  ...    & (136.97, 0.055) &(66.11, 0.055) \\ 	
4 & ... 	      &  ...    & (236.47, 0.055) &(114.15, 0.075) \\ 	
5 & ... 	      &  ... 	        & ...     &(125.02, 0.055) \\ 	
6 & ... 	      &  ... 	        & ...     &(587.44, 0.045) \\ 	
7 & ... 	      &  ... 	        & ...     &(704.72, 0.055) \\ 	
8 & ... 	      &  ... 	        & ...     &(6858.18, 0.045) \\ 	
\hline
\end{tabular}
\tablecomments{Each entry is given in ($h\nu_n, L_n$) where $h\nu_n$
is in unit of eV and the sum of $L_n$ is not necessarily equal to
one.}
\label{table:sed_mirocha}
\end{center}
\end{table}

Figure~\ref{fig:sed} shows the SED bin energies in log space chosen for
different number of energy bins: $N_\nu=$\,1\,(blue symbol), 2\,(green),
4\,(red), and 8\,(cyan). With increasing $N_\nu$, the distribution
occupies the energy space more evenly. If the SEDs are chosen in such
way that the energy under the curve is preserved, then the lowest energy
point shifts toward lower energy values with increasing $N_\nu$. Such SEDs have
a higher fraction of the ionizing UV photons in proportion to the higher
energy ones. The number of the readily absorbed, lower energy UV photons
is on the other hand a primary factor determining the temperature
profile and size of the \hii~region, as described above. Thus,
considering the energy under the SED curve is a necessary but not
sufficient condition if the goal is to obtain consistent results of
photo-ionization calculations. The number of ionizing photons assigned
to the first SED bin is also of importance. In order to achieve both
conditions, we first construct SEDs according to the ``energy under the
curve" requirement and modify them, so that the ``number of photons in
the first SED bin'' condition is satisfied.

An example is shown in Figure~\ref{fig:sed} where squares mark the
modified SED energy distribution for $N_\nu=8$ which is intended to
match the properties of the $N_\nu=4$ case for the first energy bin and
the total number of ionizing photons. For the first bin of $N_\nu=8$ we
increase the energy from 21.5\,eV to 28.4\,eV and also increase the
energy fraction from $L_1=0.4318$ to $L_1^{\prime}=0.4318\times
(28.4\,{\rm eV}/21.5\,{\rm eV})=0.5704$ which is adopted in
Table~\ref{table:sed}. The extension to an even larger numbers of SED
energy bins $N_\nu \ga 16$ should be modeled carefully since more than 2
energy bins are allotted in this case for $E< \langle E \rangle$.

Figure~\ref{fig:sed_test} illustrates the results of simulations with
different SED configurations. The SED configurations with $N_\nu=4$ (red
solid) and 8 (blue solid), in which only energy under the curve
criterion was considered, clearly exhibit different temperature of the
ionized region: $T_{\rm HII} \sim 4.5\times 10^4$\,K and $4\times
10^4$\,K, respectively. As a consequence, the accretion rate is higher
for $N_\nu=8$ case since for the Bondi accretion rate $\dot{M}_{\rm BH}
\propto T_{\rm HII}^{-3/2}$. 

In the next step, the SED with $N_\nu=8$ is modified in such way that
its first energy bin has an equal energy and number of UV photons as
$N_\nu=4$ model (M4N3E8mod, red dashed line). The two runs exhibit
indistinguishable temperature evolution in the \hii~region and nearly
the same (albeit not identical) evolution in the accretion rate. This
demonstrates that our approach to modeling the SEDs will give consistent
results regardless of the exact numerical configuration of the SED. As
an additional test of our method, we also change the first bin for
$N_\nu=4$ (blue dashed) to mirror the energy but not the number of
photons in the first bin of unmodified SED with $N_\nu=8$ (M4N3, blue
solid). As a result, the temperature evolution in the \hii~region is
indistinguishable in the two models but the evolution of the accretion
rate differs noticeably between $N_\nu=8$ (blue solid) and $N_\nu=4$
with modification (blue dashed).

We also investigate the effects of numerical resolution coupled with the
physics of photo-ionization in a run with a modified SED. M4N3E8R64mod
(red dot-dashed) shows the highest resolution test with a $64^3$ top
grid and 3 levels of refinement. Note that in this run the first burst
of accretion is slightly delayed by $\sim$\,0.1\,Myr compared to M4N3,
however the following sequence of oscillations shows consistent results
in terms of the accretion rate and the mean period of oscillation. 

Finally, we also test the \revI{SEDs for the source with a discretized
energy spectrum, which are optimized to recover analytic solutions for the
size and thermal structure of \hii~region} using the method by
\citet{Mirocha:2012}. The SEDs based on this method are listed in
Table~\ref{table:sed_mirocha} \revI{and shown as triangles in
Figure~\ref{fig:sed}}. Note that with increasing $N_\nu$, the energy
of the first bin is fixed at $E \sim 18.5\,$eV which is similar to
our strategy for selecting the first energy bin. Green solid
($N_\nu=4$) and dashed ($N_\nu=8$) lines show the results for this
SED configuration. The temperature of \hii~region is $T_{\rm HII}
\sim 30,000\,$K which is lower than other simulations, and can be
understood as a consequence of assigning lower energy to the first
SED bin.

\bibliographystyle{aasjournal}
\bibliography{park_bh}

\end{document}